\documentstyle[]{mn}
\input psfig.tex

\newif\ifAMStwofonts

\ifoldfss
  \ifCUPmtlplainloaded \else
    \NewTextAlphabet{textbfit} {cmbxti10} {}
    \NewTextAlphabet{textbfss} {cmssbx10} {}
    \NewMathAlphabet{mathbfit} {cmbxti10} {} % for math mode
    \NewMathAlphabet{mathbfss} {cmssbx10} {} %  "   "    "
  \fi
  \ifAMStwofonts
    \ifCUPmtlplainloaded \else
      \NewSymbolFont{upmath} {eurm10}
      \NewSymbolFont{AMSa} {msam10}
      \NewMathSymbol{\upi}     {0}{upmath}{19}
      \NewMathSymbol{\umu}     {0}{upmath}{16}
      \NewMathSymbol{\upartial}{0}{upmath}{40}
      \NewMathSymbol{\leqslant}{3}{AMSa}{36}
      \NewMathSymbol{\geqslant}{3}{AMSa}{3E}

    \fi
  \fi
\fi % End of OFSS

\ifnfssone
  \newmathalphabet{\mathit}
  \addtoversion{normal}{\mathit}{cmr}{m}{it}
  \addtoversion{bold}{\mathit}{cmr}{bx}{it}
  \newmathalphabet{\mathbfit} % math mode version of \textbfit{..}
  \addtoversion{normal}{\mathbfit}{cmr}{bx}{it}
  \addtoversion{bold}{\mathbfit}{cmr}{bx}{it}
  \newmathalphabet{\mathbfss} % math mode version of \textbfss{..}
  \addtoversion{normal}{\mathbfss}{cmss}{bx}{n}
  \addtoversion{bold}{\mathbfss}{cmss}{bx}{n}
  \ifAMStwofonts
    \ifCUPmtlplainloaded \else
      \UseAMStwoboldmath
      \makeatletter
      \new@mathgroup\upmath@group
      \define@mathgroup\mv@normal\upmath@group{eur}{m}{n}
      \define@mathgroup\mv@bold\upmath@group{eur}{b}{n}
      \edef\UPM{\hexnumber\upmath@group}
      \new@mathgroup\amsa@group
      \define@mathgroup\mv@normal\amsa@group{msa}{m}{n}
      \define@mathgroup\mv@bold\amsa@group{msa}{m}{n}
      \edef\AMSa{\hexnumber\amsa@group}
      \makeatother
      \mathchardef\upi="0\UPM19
      \mathchardef\umu="0\UPM16
      \mathchardef\upartial="0\UPM40
      \mathchardef\leqslant="3\AMSa36
      \mathchardef\geqslant="3\AMSa3E
    \fi
  \fi
\fi % End of NFSS release 1

\ifnfsstwo
  \DeclareMathAlphabet{\mathbfit}{OT1}{cmr}{bx}{it}
  \SetMathAlphabet\mathbfit{bold}{OT1}{cmr}{bx}{it}
  \DeclareMathAlphabet{\mathbfss}{OT1}{cmss}{bx}{n}
  \SetMathAlphabet\mathbfss{bold}{OT1}{cmss}{bx}{n}
  \ifAMStwofonts
    \ifCUPmtlplainloaded \else
      \DeclareSymbolFont{UPM}{U}{eur}{m}{n}
      \SetSymbolFont{UPM}{bold}{U}{eur}{b}{n}
      \DeclareSymbolFont{AMSa}{U}{msa}{m}{n}
      \DeclareMathSymbol{\upi}{0}{UPM}{"19}
      \DeclareMathSymbol{\umu}{0}{UPM}{"16}
      \DeclareMathSymbol{\upartial}{0}{UPM}{"40}
      \DeclareMathSymbol{\leqslant}{3}{AMSa}{"36}
      \DeclareMathSymbol{\geqslant}{3}{AMSa}{"3E}
    \fi
  \fi
\fi % End of NFSS release 2

\ifCUPmtlplainloaded \else
  \ifAMStwofonts \else % If no AMS fonts
    \def\upi{\pi}
    \def\umu{\mu}
    \def\upartial{\partial}
  \fi
\fi

\def\simlt{\lower.5ex\hbox{$\; \buildrel < \over \sim \;$}}
\def\simgt{\lower.5ex\hbox{$\; \buildrel > \over \sim \;$}}
\def\ms{M$_{\odot}$}

\def\mp{M$_{\odot}$ pc$^{-2}$}

\def\l{$\lambda$ }

\begin{document}

\title{Chemo-spectrophotometric evolution of spiral galaxies: \\
     II. Main properties of present day disk galaxies}

\author[S. Boissier and N. Prantzos]
       {S. Boissier and N. Prantzos \\
 Institut d'Astrophysique de Paris, 98bis, Bd. Arago, 75014 Paris}
\date{ }

\pagerange{\pageref{firstpage}--\pageref{lastpage}}
\pubyear{1999}
\maketitle

\label{firstpage}

\begin{abstract}

We study the chemical and spectro-photometric evolution of galactic disks
with detailed models calibrated on the Milky Way and using simple
scaling relations, based on currently popular semi-analytic models
of galaxy formation. We compare our results to a large body of observational
data on present day galactic disks, including:  
disk sizes and central surface brightness, Tully-Fisher relations in various 
wavelength bands, colour-colour and colour-magnitude relations, gas fractions 
vs. magnitudes and colours, abundances vs. local and integrated properties,
as well as spectra for different galactic rotational velocities. 
Despite the extremely simple nature of our models, we find satisfactory
agreement with all those observables, provided the timescale for star
formation in low mass disks is longer than for more massive ones.
This assumption is apparently in contradiction with the standard picture of
hierarchical cosmology. We find, however, that it is extremely successfull
in reproducing major features of present day disks, like the change in the 
slope of the Tully-Fisher relation with wavelength, 
the fact that more massive galaxies
are on average ``redder'' than low mass ones  (a generic problem of standard
hierarchical models) and  the metallicity-luminosity relation for spirals.
It is concluded that, on a purely empirical basis, this new picture at least as
successful as the standard one. Observations at high redshifts could help
to distinguish between the two possibilities.

\end{abstract}


\section{Introduction}

Considerable progress has been made in the past ten years or so towards a
qualitative understanding of galaxy properties. This has been made in the
framework of the Cold Dark Matter (CDM) picture for galaxy formation, either
with full N-body simulations coupled to gas hydrodynamics 
(e.g. Steinmetz 1998) or with simpler,
so-called semi-analytic (SAM) models (e.g. Somerville and Primack 1998).

N-body simulations have difficulties up to now in reproducing the observed
properties of galaxies in detail. In particular they  fail to produce
realistic models for present day spirals, because excessive transfer of 
angular momentum from gas to dark matter results in overly small disks (e.g.
Navarro and Steinmetz 1997). Moreover, the first attempt to combine in a single
computation dynamical, chemical and photometric evolution (Contardo et al.
1998) failed to reproduce basic observational properties of the Milky Way disk
(like the local G-dwarf metallicity distribution or colour gradients at low
redshift).

Semi-analytic models, pioneered by White and Frenk (1991) and subsequently
augmented by Monte-Carlo methods, were developed by several groups
(``Munich'': Kauffmann, White and Guiderdoni 1993, Kauffmann 1996, Kauffmann, 
Nusser and Steinmetz  1997; ``Durham'': Cole et al. 1994, Baugh et al. 1997;
``Santa-Cruz'': Somerville and Primack 1998). Such models include a 
simplified treatment of several essentially unknown ingredients, like
star formation, gas cooling, supernova feedback and galaxy merging. 
They constitute an efficient tool to explore the large parameter space
occupied by these (and several other!) unknowns, but they also
provide an important level of understanding that could not be achieved by
running N-body simulations alone. The successes and shortcomings of such
models are nicely reviewed in Somerville and Primack (1998).

Despite the apparent  success of SAMs, it is not  obvious  whether
their number of free parameters is smaller than the number of the observables
they successfully explain. One rarely mentionned shortcoming is that the
assumption of Instantaneous Recycling Approximation (IRA) results in 
overestimated metallicities and underestimated gas fractions at late times.
Moreover, the wealth of available spatial information (i.e. data
on gradients of colours, gas and metallicity) as rarely used as constraints 
to SAMs (with the exception of Kaufmann 1996).

This latter point is dealt with in Jimenez et al. (1998), who use the simple
scaling relations for galactic disks (established in the framework of SAMs, see
Sec. 2.2) in order to derive initial gaseous profiles. They subsequently
calculate the full chemical and spectrophotometric evolution of such disks
for the case of Low Surface Brightness (LSB) galaxies and 
compare succesfully their 
resulting colour and abundance gradients to observations.

In this paper we adopt an approach similar to that of Jimenez et al. (1998), 
i.e. we use multi-zone  models and simple scaling laws given by SAMs
in order to study the evolution of galactic disks. Our method and purpose are,
however, different. We calibrate our models to a fairly detailed ``template'' 
of the Milky Way (elaborated in a previous paper: Boissier and Prantzos 1999,
paper I in this series), while Jimenez et al. (1998) make no such calibration.
Moreover, we use a different prescription for the radial dependence of the
star formation rate, inspired by theories of star formation in spiral disks
(see Sec. 2.1).
Finally, we are interested mainly on the properties of High Surface Brightness
disks, although (as we shall see) some of our models produce LSBs and some
bulges or ellipticals.

Our main objective is to check whether in such an oversimplified framework
(i.e. neglecting all the poorly known physics of star-gas interactions, galaxy
merging  etc.)
one can reproduce  observations concerning present day disks. For that purpose
we use a large body of observational data on disks, concerning
sizes and central surface brightness, Tully-Fisher relations in various 
wavelength bands, colour-colour and colour-magnitude relations, gas fractions 
vs. magnitudes and colours, abundances vs. local and integrated properties,
as well as spectra for different rotational velocities. Comparison of the
models to the data turns out to be successfull, provided a crucial assumption
is made: that small galaxies form on average their stars {\it later}
than their more massive counterparts. Hierarchical models for galaxy
formation produce a radically different picture: low mass objects are
generally predicted to form {\it earlier} than more massive ones and 
consequently
produce older and ``redder'' stellar populations, contrary to observations.
In Somerville and Primack (1998), dust extinction, assumed to be more important
in more massive galaxies, is invoked in order to reverse the trend and 
bring agreement between models and  observations.
In our models dust extinction plays a minor role, merely enhancing an already
correct trend, while at the same time several other observables 
 are naturally reproduced.

The plan of the paper is as follows: In Sec. 2 we present the basic 
ingredients and the underlying assumptions of the model. The main input 
parameters and an application to the Milky Way (already discussed in detail
in Paper I) are presented in Sec. 2.1. The extension to other galactic disks
with the help of simple scaling relations (established in the framework of 
CDM models) is presented in Sec. 2.2. The assumptions about the infall and
star formation rate are discussed in more details in Sec. 2.3.
Some results of  the models (evolution of the gaseous and stellar content,
star formation rate, colour profiles etc.), as well as the main present
day properties, are presented in Sec. 3.  
The results are compared to observations in Sec. 4:
disk sizes and central surface brigthness (Sec. 4.1), 
Tully-Fisher relations in various  wavelength bands (Sec. 4.2), 
colour-colour and colour-magnitude relations (Sec. 4.3), 
gas fractions  vs. magnitudes and colours (Sec. 4.4), 
abundances vs. local and integrated properties (Sec. 4.5),
integrated  spectra for different galactic rotational velocities (Sec. 4.6).
The results are summarised in Sec. 5, where some comments are also made
on the emerging overall picture.

\section{The model}

In the previous paper of this series (BP99) we presented a detailed model
of the chemical and spectrophotometric evolution of a spiral galaxy and 
we applied it to the case of the Milky Way disk. In this section, we recall
briefly the main ingredients of the model and the results for our Galaxy
(Sec. 2.1). Assuming that the Milky Way is a typical spiral galaxy, we
extend the model to the study of other spirals, helped by some
simple ``scaling laws'' (Sec. 2.2), which have been established in the framework
of the Cold Dark Matter scenarios for galaxy formation.
The SFR is then calculated in a self-consistent way, while a further ``scaling law''
has to be assumed for the infall rate, a crucial ingredient in all realistic
models of spiral evolution (see Sec. 2.3).
 
\subsection{Main ingredients and application to the Milky Way}

 The galactic disk is simulated as an ensemble of concentric, independently
evolving rings, gradually built up by infall of primordial composition. The
chemical evolution of each zone is followed by solving the appropriate
set of integro-differential equations, without the Instantaneous Recycling
Approximation. Stellar yields are from Woosley and Weaver (1995) for massive stars
and Renzini and Voli (1981) for intermediate mass stars. Fe producing SNIa are
included, their rate being calculated with the prescription of Matteucci and
Greggio (1986). The adopted stellar Initial Mass Function (IMF)
is a multi-slope power-law between 0.1 \ms \ and 100 \ms \ from the work of
Kroupa et al. (1993).

The star formation rate (SFR) is locally given by a
Schmidt-type law, i.e. proportional to some power of the gas surface density 
$\Sigma_g$: $\Psi \propto \Sigma_g^{1.5}$, according to the
observations of Kennicutt (1998); it varies with galactocentric radius $R$, as:
\begin{equation}
 \Psi(t,R) \ = \  \alpha \  \Sigma_g(t,R)^{1.5} \ V(R) \ R^{-1}
\end{equation}
where $V(R)$ is the circular velocity at radius $R$. This radial dependence of
the SFR is suggested by
the theory of star formation induced by density waves in spiral
galaxies (e.g. Wyse and Silk 1989). Since $V(R)\sim$const. in the largest part 
of the disk, this is equivalent to $\Psi(R) \propto \Sigma_g(R)^{1.5} R^{-1}$.
This latter form of the SFR is used in Prantzos and Silk (1998) and BP99. 
However, when extending the model
to other galaxies with different rotational velocities, we use the form of
Eq. (1), and we drop the assumption of $V(R)=const.$ by calculating the full
rotation curve (see Sec. 2.3).

The infall rate is assumed to be exponentially decreasing in time, i.e.
\begin{equation}
f(t,R) \  = \ A(R) \ e^{-t/\tau(R)}
\end{equation}
with $\tau(R_0$=8 kpc) = 7 Gyr, in order to reproduce the local G-dwarf
metallicity distribution and $\tau(R)$ increasing outwards, from 
$\tau(R$=2 kpc)=1 Gyr to $\tau(R$=17 kpc)=12 Gyr
(see Fig. 3).  This radial dependence  of
$f(R)$ is simulating the inside-out formation of galactic disks and, combined
with the adopted $\Psi(R)$ allows to reproduce the observed current profiles
of gas, oxygen abundance and SFR in the Milky Way disk (see BP99). The coefficient
$A(R)$ is obtained by the requirement that at time T=13.5 Gyr, the current
mass profile of the disk $\Sigma(R)$ is obtained, i.e.
\begin{equation}
\int_0^T f(t,R) \  = \ \Sigma(R)
\end{equation}
with $\Sigma(R) \ \propto e^{-R/R_G}$ and a scalelength $R_G$=2.6 kpc for the
Milky Way disk.

The spectrophotometric evolution is followed in a self-consistent way, i.e.
with the  SFR $\Psi(t)$ and metallicity $Z(t)$ of every zone determined by the chemical evolution,
and the same IMF. The stellar lifetimes, evolutionary tracks and spectra are
metallicity dependent; the first two are from the Geneva library (Schaller  et al. 1992,
Charbonnel et al. 1996) and the latter from Lejeune et al (1997). Dust absorption is
included according to the prescriptions of  
Guiderdoni et al. \shortcite{guiderdoni98}
and assuming a ``sandwich''
configuration for the stars and dust layers (Calzetti et al. 1994).

This model has several ingredients (stellar yields, tracks, lifetimes, spectra)
calculated by the (presumably well understood) theories of stellar atmospheres,
evolution and nucleosynthesis, while it has relatively few free parameters.
Indeed, we assume that the IMF (obtained by local observations with particular
care to the evaluation of the low mass part by  Kroupa et al. 1993) 
and the dependence of the SFR on gas
surface density (derived from observations of external spirals, Kennicutt 1998)  
are not free  parameters. On the other hand, the efficiency  $\alpha$
of the SFR  (Eq. 1) is fixed
by the requirement that the local gas fraction
$\sigma_g(R_0$=8 kpc)$\sim$0.2,
is reproduced at T=13.5 Gyr. We consider then that the really ``free'' parameters
of the model are the radial dependence of the
infall timescale $\tau(R)$ and of the SFR; in the latter case, the radial dependence
(the $V(R)/R$ factor) has some theoretical motivations (see Wyse and Silk 1989).

The number of observables explained by the model is much
larger than the number of free parameters. Leaving aside local observables
(age-metallicity relationship, G-dwarf metallicity distribution, O/Fe evolution)
the model reproduces present day ``global'' ones 
(profiles of gas, O/H, SFR, as well as supernova rates).
Moreover, the current disk luminosities in various wavelength bands are
successfully obtained along with the corresponding radial profiles; in particular,
the adopted inside-out star forming scheme leads to a scalelength of $\sim$4 kpc
in the B-band and $\sim$2.6 kpc in the K-band, in agreement with observations
(see BP99 for references and details).
It is the first time, to our knowledge, that a consistent model for the chemical
{\it and} photometric evolution of the Milky Way is developed. Its success encourages us
to extend it to other disk galaxies, in the way described below.

\subsection{Scaling properties of disk galaxies}

For a simplified description of disk galaxies we adopt here the ``scaling properties''
derived by Mo, Mao and White (1998, hereafter MMW98) in the framework of
the Cold Dark Matter (CDM) scenario for galaxy formation. According to this
scenario, primordial density fluctuations give rise to haloes of non-baryonic dark
matter of mass $M$, within which baryonic gas condenses later and forms disks
of maximum circular velicity $V_C$. The mass of the haloes is
\begin{equation}
M \ = \ \frac{V_C^3}{10 \ G \ H(z)}
\end{equation}
where $H(z)$ is the Hubble parameter at the redshift $z$ of halo formation and
$G$ the gravitational constant.
The mass of the disk $M_d$ is a fraction $m_d$ of the halo mass
\begin{equation}
M_d \  = \ m_d \ M
\end{equation}
It is assumed that the disk is thin, rotationally supported and has an exponential
surface density profile
\begin{equation}
\Sigma(R) \ = \ \Sigma_0 \ e^{-R/R_d}
\end{equation}
with central density $\Sigma_0$,    scalelength $R_d$
and corresponding mass
\begin{equation}
M_d  \ = \ 2 \ \pi \ \Sigma_0 \ R_d^2
\end{equation}

Eq. (6) describes the initial  gaseous  profile of the disk; in the terms of
our infall model of Sec. 2.1, $\Sigma(R)$ corresponds to the right hand side
of Eq. (3). Eqs. (4), (5) and (6) allow to relate the parameters $\Sigma_0$ and
$R_d$ to an observable, the circular velocity of the disk $V_C$, but not in a
unique way (disks of the same mass $M_d$ and circular velocity 
$V_C$ may have different $\Sigma_0$ and
$R_d$). In order to break the degeneracy one may adopt
the spin parameter $\lambda$ as a second parameter, as e.g.
in MMW98. This parameter is
 related to the halo mass $M$ and angular momentum $J$ through
\begin{equation}
\lambda \ = J \ E^{1/2} \ G^{-1} \ M^{-5/2}
\end{equation}
where the total energy $E$ of the halo (assumed to be an isothermal sphere)
is given by
\begin{equation}
E \ = \  -\frac{M \ V_C^2}{2}
\end{equation}
Finally, assuming that the angular momentum of the disk
\begin{equation}
J_d \  = \ 2 \ M_d \ V_c \ R_d
\end{equation}
is a fraction $j_d$ of that of the halo, i.e.
\begin{equation}
J_d \  = j_d \ J
\end{equation}
one may express the disk parameters $\Sigma_0$ and $R_d$ in terms of
$V_C$ and $\lambda$, as well as $j_d$, $m_d$ and $H(z)$:
\begin{equation}
R_d \  = \  \frac{1}{10 \ \sqrt{2}} \ \lambda \ V_C \
\left(\frac{j_d}{m_d}\right) \ H(z)^{-1}
\end{equation}
and
\begin{equation} % DANS CETTE EQUA, J'AI REMONTE LE 10 .....................
\Sigma_0 \  = \ \frac{10}{ \ \pi \ G} \
m_d \ \lambda^{-2} \ V_C \ \left(\frac{j_d}{m_d}\right)^{-2}
 \ H(z)
\end{equation}

\begin{figure}
 \psfig{file=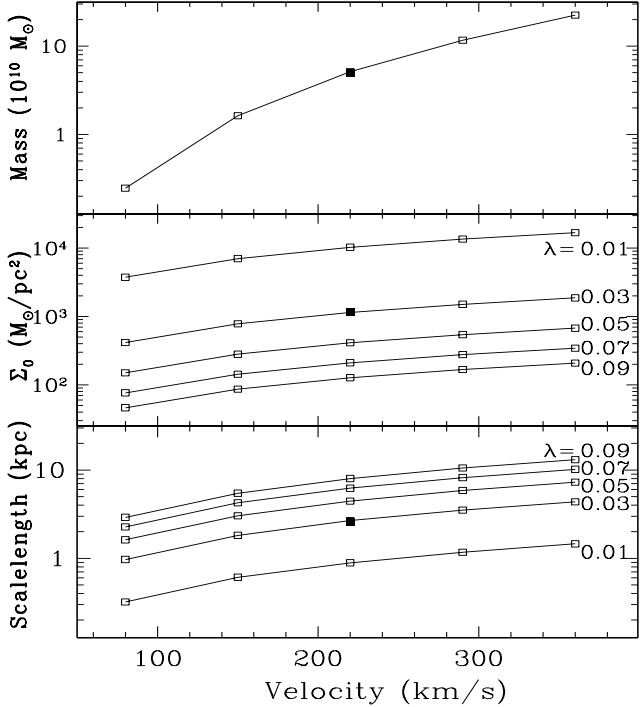,height=10.cm,width=0.5\textwidth}
\caption{\label{FIG1} {Main properties of our model disks. From top to
bottom: mass, central surface density and scalelength, respectively.
They are plotted as a function of circular velocity $V_C$ and 
parametrised with the spin parameter $\lambda$. 
Disk mass depends only on $V_C$. Filled symbols correspond
to the Milky Way model, used for the scaling of all other disks
(through eq. 14 and 15, with $m_d=m_{dG}$=0.05).
}}
\end{figure}

In this work we ignore the effects of the formation time of the disk,
entering Eqs. (12) and (13) through the term $H(z)$.
Indeed, it is difficult to incorporate it in the framework of our infall model
since the disk is slowly built up and the term ``time of formation'' has not
a precise meaning. We will assume then that all disk galaxies 
{\it start forming their stars  at the same
time (or redshift) and have today the same age as the Milky Way, i.e.  T=13.5 Gyr} 
(notice that the precise
formation time is not very important, as far as the corresponding redshift is
$z > 3$).
We will discuss, however, later the
possibility that present day disks have different ages, which is equivalent to
assuming different redshifts for their formation.

It is usually assumed that $m_d$ and $j_d$ are constants and, moreover, that
$j_d/m_d$=1, i.e. that the specific angular momentum of the material that
forms the disk and the dark halo are the same
(see MMW98 and references therein). The latter assumption, although not
supported by  numerical simulations,
simplifies considerably our modelling (otherwise, radial inflows should be induced
by angular momentum transfer in the disk, invalidating our model of independently
evolving rings).
In that case, the profile of a given disk can be expressed in terms of the one of our
Galaxy (the parameters of which are designated hereafter by index G):
\begin{equation}
\frac{R_d}{R_{dG}}  \  = \  \frac{\lambda}{\lambda_G} \ \frac{V_C}{V_{CG}}
\end{equation}
and
\begin{equation}
\frac{\Sigma_0}{\Sigma_{0G}}  \  = \  \frac{m_d}{m_{dG}} \
 \left(\frac{\lambda}{\lambda_G}\right)^{-2}
 \ \frac{V_C}{V_{CG}}
\end{equation}
where we have: $R_{dG}$=2.6 kpc, $\Sigma_{0G}$=1150 \mp, $V_{CG}$=220 
km/s (see BP99 for references on the properties of the Milky Way).
Notice that the assumption that all galaxies started forming at the same
redshift eliminated $H(z)$ from Eq. 14 and 15.
The mass of the disk is $M_{dG}$=5 10$^{10}$ \ms \ and, assuming
$m_d$=0.05 we obtain M$_G$=10$^{12}$ \ms \ for the mass of the
Milky Way's dark  halo. Using equs. 5, 8, 9, 10 and 11 one can also
evaluate the spin parameter $\lambda$ as:
\begin{equation}
\lambda \ = \ \left(\frac{\sqrt{2}}{G}\right) \
\left(\frac{m_d}{j_d}\right) \ R_d \ V_c^2 \ M^{-1}
\end{equation}
Using the above numerical values for the Milky Way (and $m_d/j_d$=1), 
we find $\lambda_G$=0.03.
Eqs. 14 and 15  allow  to describe the mass profile of a galactic disk
in terms of the one of our Galaxy and of two parameters: $V_C$ and $\lambda$.
The range of observed values for the former parameter
is 80-360 km/s, where for the latter
numerical simulations give values in the 0.01-0.1 range, the distribution
peaking around $\lambda\sim0.04$ (MMW98). % Autre ref ? Cole-Lacey ? 
Although it is not clear yet whether
$V_C$ and $\lambda$ are independent quantities, we treat them here as such
and construct a grid of 25 models caracterised by $V_C$ = 80, 150, 220, 290, 360 km/s
and $\lambda$ = 0.01, 0.03, 0.05, 0.07, 0.09, respectively.
The corresponding main properties of the disks (mass $M_d$,
central surface density $\Sigma_0$ and disk scalelength $R_d$)
appear in Fig. 1, expressed
in terms of the adopted parameters $V_C$ and $\lambda$.
Larger values of $V_C$ correspond to more massive disks and
larger values of $\lambda$ to more extended ones.

As we shall see in Sec. 4.1 the resulting disk radii and central
surface brightness are in excellent agreement with observations, 
except for the case of $\lambda$=0.01.
We could have excluded that value from our grid on the basis of
instability arguments. Indeed, disks dominated by their self-gravity are
likely to be unstable to the formation of a bar. According to the
discussion in MMW98, for $m_d$=0.05, this would happen for disks
with $\lambda<$0.03 (see their Fig. 3). We prefer, however, to keep this
$\lambda$ value in our grid, for illustration purposes; as we shall see,
this ``unphysical'' value leads to the production of galaxies ressembling
to bulges or ellipticals, rather than disks.

\begin{figure}
\psfig{file=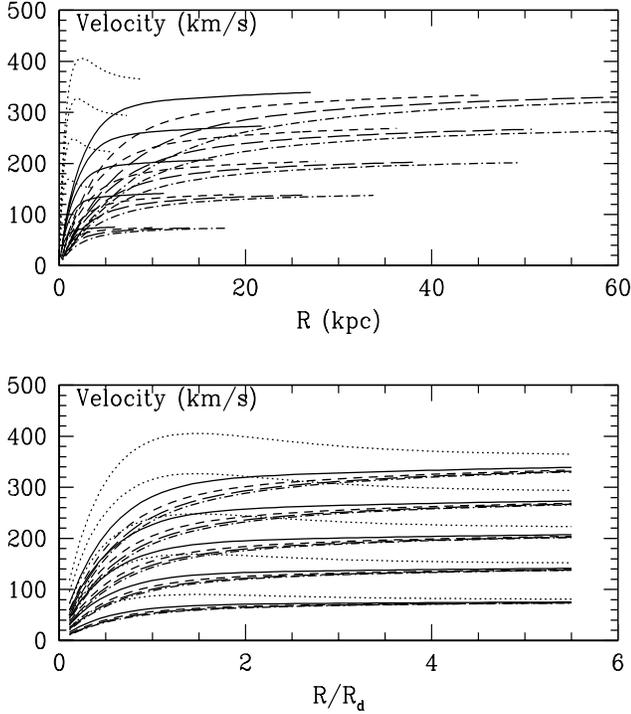,height=10.cm,width=0.5\textwidth}
\caption{\label{FIG2} {Rotation curves of our disk models as function of
radius, expressed in kpc ({\it upper pannel}) and in disk scalelengths $R_d$
({\it lower pannel}). For each of the five values of our grid $V_C$ (80, 150, 220,
290 and 360 km/s) we plot the results for the corersponding five $\lambda$ 
values ({\it dotted} for 0.01, {\it solid} for 0.03, {\it short-dashed} for
0.05, {\it long-dashed} for 0.07 and {\it dot-dashed} for 0.09).
}}
\end{figure}

\subsection{SFR and Infall}

The two main ingredients of
the model, namely the Star Formation Rate $\Psi(R)$ and the infall time-scale
$\tau(R)$, are affected by the adopted scaling of disk properties
in the following way.

For the SFR we adopt the prescription of Eq. (1), with the same efficiency $\alpha$
as in the case of the Milky Way.
In order to have an accurate evaluation of
$V(R)$ across the disk, we calculate it as the sum of the contributions of the disk
(with the surface density profile of Eq.  6) and of the dark halo, with a volume
density profile given by:
\begin{equation}
\rho(R) \ = \  \frac{\rho_0}{1+(\frac{R}{R_C})^2}
\end{equation}
where $\rho_0$ is the central density and $R_C$ the core radius 
(profile on a non-singular isothermal sphere).
Other halo profiles (e.g. Navarro et al. 1997) are, probably, more
realistic, but their effect on our results is negligible
(the resulting rotation curves differ by less than 20\% from the one adopted here
except in the innermost regions, where  spiral waves are less effective and the
presence of bulges and bars makes the SFR very uncertain, anyway).
The corresponding contributions of the disk and the halo to the
rotational velocity are given by (Navarro 1998):
\begin{equation}
V_d(R) \ = \ 2 \ \pi \ G \ \Sigma_0 \ R_d \ x \ [I_0(x)K0(x)-I1(x)K1(x)]^{\frac{1}{2}}
\end{equation}
where $x=R/R_d$ and $I_n$ and $K_n$ are modified Bessel functions of order n ; and
\begin{equation}
V_h(R) \ = \ 4 \ \pi \ G \ \rho_0 \ R_C^2 [1- \frac{R_C}{R} arctan (\frac{R}{R_C})]
\end{equation}
respectively, while the final rotation curve is calculated by:
\begin{equation}
V(R)^2 \ = \ V_h(R)^2 \ + \ V_d(R)^2
\end{equation}
The calculated rotation curves for our models appear in Fig. 2, as
 a function of radius expressed in kpc ({\it upper pannel}) and in scalelengths 
$R_d$ ({\it lower pannel}); it can
be seen that in the latter case, a kind of ``universal'' rotation curve
is obtained (excpet for the unrealistic, as we shall see below, $\lambda$=0.01 case). 
Although this velocity profile is not exactly the same as the one derived
observationally in Salucci and Persic (1997), it is very close to it and
certainly  sufficient for the purposes of this work. It allows to calculate
the SFR  (Eq. 1) in a fully self-consistent way.

The choice of the infall rate $\tau(R)$ is more problematic than 
the one of the SFR because
there are no theoretical principles     to guide us.
Molla et al. \shortcite{molla98} adopt
an exponential radial dependence (with a scale-length
of 4 kpc) for $\tau(R)$ in the Milky Way disk and a characteristic
time $\tau_0=\tau_{\odot} (M_d/M_{dG})^{-1/2}$ for the other galactic disks;
this scaling is made on the basis of the association $R_{eff}/R_0$
(effective radius/solar galactocentric distance) between the Milky Way and
the   galaxies of their study.

\begin{figure}
\psfig{file=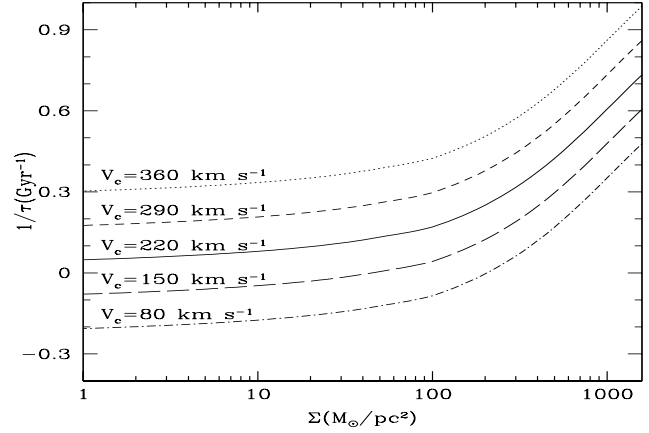,height=6.cm,width=0.5\textwidth}
\caption{\label{FIGINF} {Adopted relation between infall time scale $\tau$
and total surface density $\Sigma$,
generalized from the Milky Way model ({\it solid curve}) to other galaxies
(see Sec. 2.3). Infall is more rapid in denser regions and in more massive galaxies.
Negative values of $1/\tau$ mean that the infall rate is increasing in time.
}}
\end{figure}

In our case, we have no a priori idea on which region of a given disk will be
equivalent to the solar neighborhood.
We then adopt the following prescription:
the infall time scale increases with both surface density (i.e.
the denser inner zones are formed more rapidly) and with galaxy's mass,
i.e. $\tau[M_d,\Sigma(R)]$.  
In both
cases  it is the larger gravitational potential that induces a more rapid infall.
Our adopted infall rate is given by:
\begin{equation}
f[M_d,\Sigma(R),t] \,  \propto \, exp \left(- \frac{t}{\tau_G(\Sigma)}+0.4 t 
\left(1-\frac{V_C(M_d)}{220}\right)\right)
\end{equation}

The corresponding characteristic infall timescales  $\tau[M_d,\Sigma(R)]$ appear 
in Fig. 3, as a function of $\Sigma(R)$ and of $V_C$.
 The  radial dependence of $\tau$ on $\Sigma(R)$ is calibrated on the Milky Way
(figure \ref{FIGINF}) where $\tau$ is fixed to 7 Gyr at $R_0$=8 kpc
(to solve the G-dwarf problem) and to 1 Gyr at  2kpc from the galactic center.
On the other hand, the mass dependence  of $\tau$ is adjusted as to reproduce
several of the properties of the galactic disks that will be discussed in Sec. 4.
 The adopted prescription allows to keep some simple scaling relations for the
infall in our models and the number of free parameters as
small as possible. We recognise that a different parametrisation could be made
with  a  - certainly - non negligible impact on the results. 
In fact, by suppressing the dependence of $\tau$ on galaxy mass
(i.e. by assuming the same $\tau[\Sigma(R)]$ relation for all galaxies)
we found   that several properties of disks - in particular, the colour magnitude
and metallicity-luminosity relations - are not well reproduced.
The fact that the adopted  simple
prescription provides a satisfactory agreement with several observed relationships
in spirals (see Sec. 4) makes us feel that it could be ultimately justified by theory
or numerical simulations of disk formation.

\begin{figure*}
\psfig{file=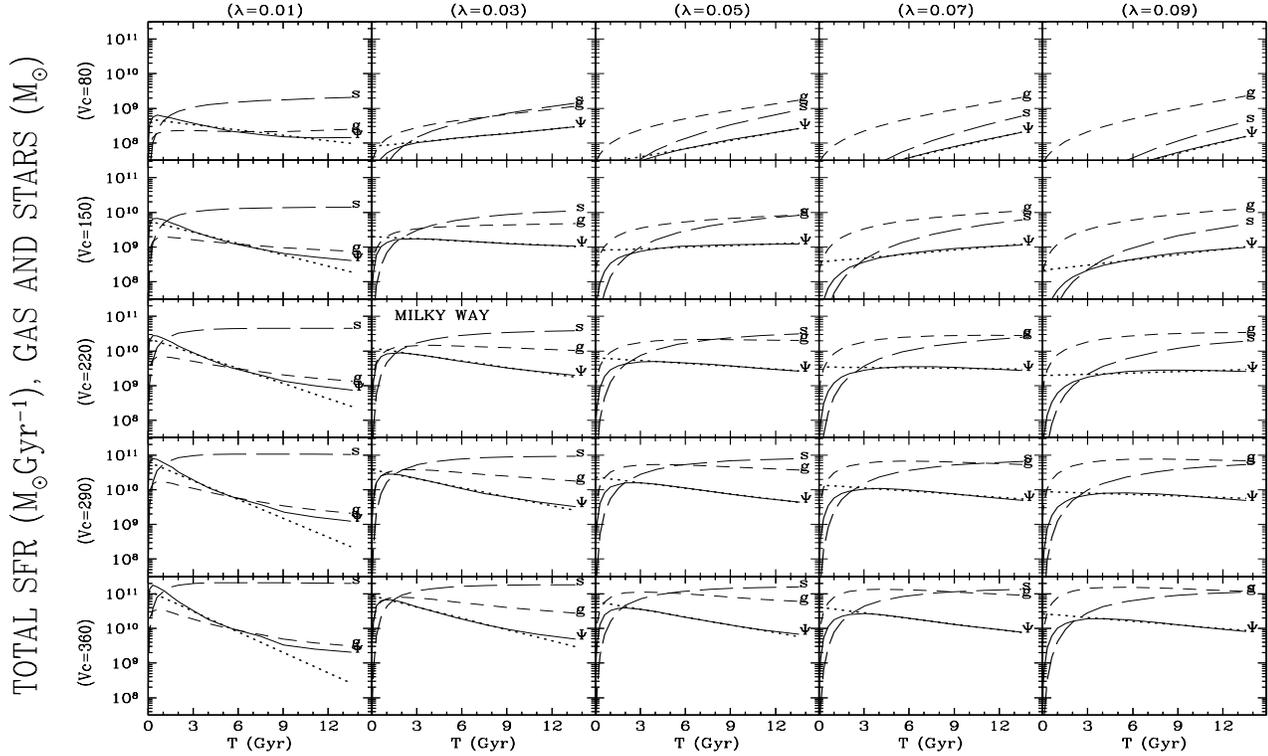,height=11.cm,width=\textwidth,angle=-90}
\caption{\label{HISTOIRE} Histories of
total SFR ($\bf \Psi$, {\it solide curve}), gas mass ($\bf g$, {\it short dashed curve})
 and stellar mass ($\bf s$, {\it long dashed curve}) for our disk
models. Each column
corresponds to a value of \l and each row to a value of $V_C$.
The {\it dotted} curve is an exponential fit to the SFR.}
\end{figure*}

\section{Model results}

We have computed the chemical and specrophotometric evolution of
a grid of 25 disk models (defined by the values  of $V_C$ and $\lambda$
given in Sec. 2.2) up to an age of T=13.5 Gyr.
We calculated the evolution of the various chemical and photometric profiles,
as well as the disk integrated properties according to the procedure outlined
in paper I (BP99, Sec. 3.3).

In this work, we concentrate mainly on the integrated properties at T=13.5 Gyr and
compare them to observations of local spirals (i.e. at redshift $z\sim$0, see Sec. 4).
However, in this section, we also present some of the results concerning 
photometric and colour profiles, since their analysis helps to understand
the behaviour of the integrated properties and the role of the adopted prescriptions
for the SFR and the infall rate. The
implications of our models for galaxy observations at higher redshifts, and the
corresponding chemical and photometric gradients will be analysed in  forthcoming
papers.

\subsection{Histories of gas, stars and SFR}

The evolution of the SFR and of the gaseous and stellar masses in our 25 disk models
is shown in Fig. 4. This evolution  results from the prescriptions for infall and
SFR adopted in Sec. 2.3. 
Values of SFR range from $\sim$100 $M_{\odot}$/yr in the early phases of massive disks
to $\sim$0.1 $M_{\odot}$/yr for the lowest mass galaxies.
The resulting SFR history is particularly interesting
when compared to simple models of photometric evolution, that are usually 
applied in studies
of galaxy evolution and of their cosmological implications. Such models (one-zone,
no chemical evolution considered in general) adopt exponentially declining
SFR with different caracteristic timescales for each galaxy morphological type 
(Guiderdoni \& Rocca-Volmerange 1987, Bruzual \& Charlot 1993, 
Fioc \& Rocca-Volmerange\shortcite{fioc97}).

\begin{figure}
\psfig{file=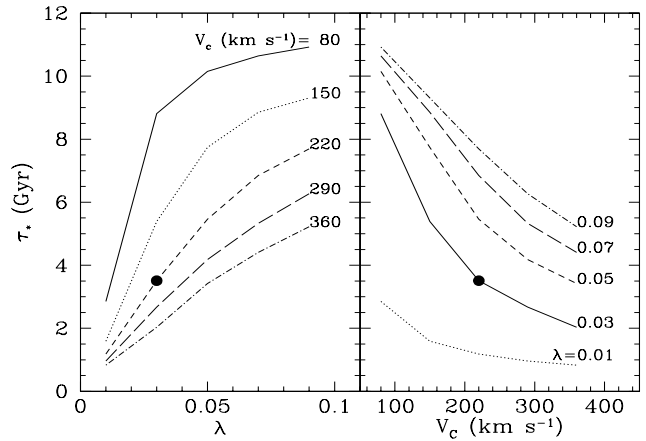,height=6.cm,width=0.5\textwidth,angle=-90}
\caption{\label{profilSurf}  Caracteristic time-scales for star formation, i.e.
for forming the first half of the stars of a given galaxy, 
resulting from our models. They
are plotted as a function of  the spin parameter $\lambda$ ({\it left}) and of
circular velocity $V_C$ ({\it right}). {\it Filled points} correspond to the Milky Way.
}
\end{figure}

\begin{figure*}
\psfig{file=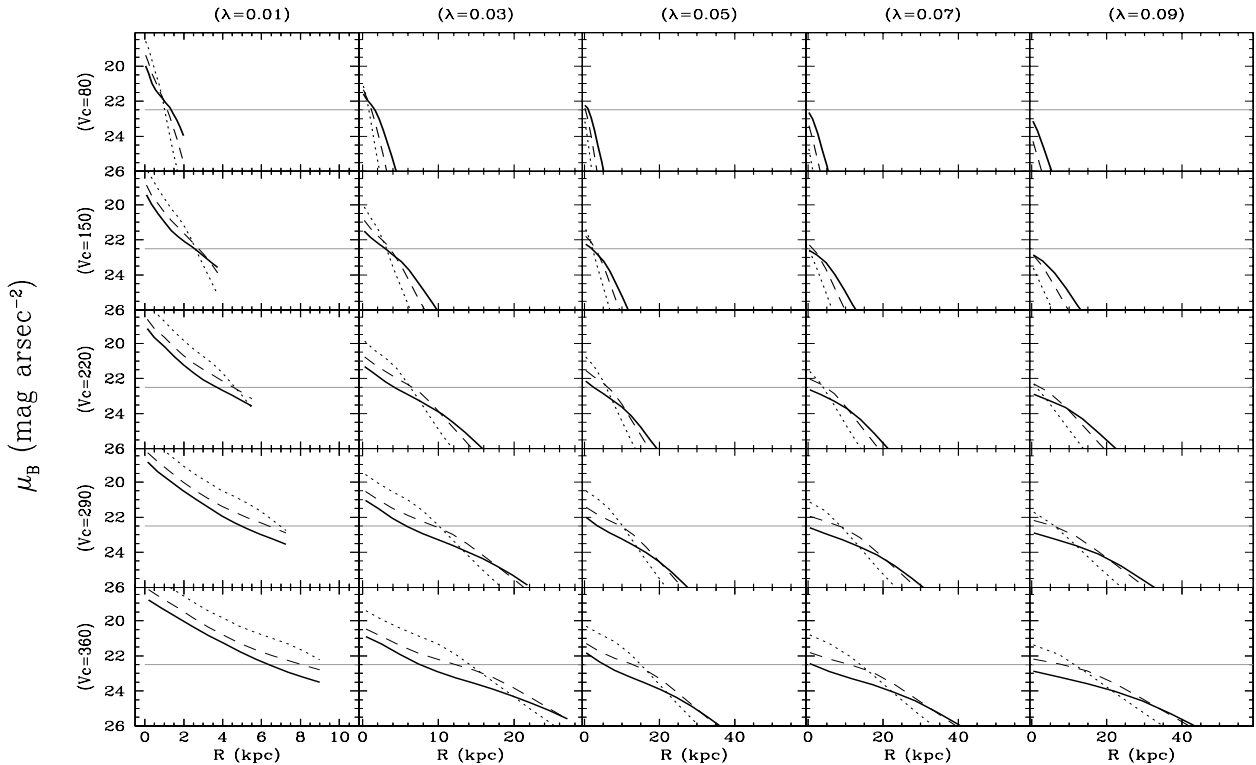,height=11.cm,width=\textwidth,angle=-90}
\caption{\label{profilB}  Evolution  of the B-band surface brightness profile of our
disk models. Profiles are shown at ages T=3 ({\it dotted}), 7.5 ({\it dashed})
 and 13.5 Gyr ({\it solid}). The {\it grey} horizontal line at $\mu_B$=22.5
mag arcsec$^{-2}$ indicates the limit for Low Surface Brightness galaxies.
Each column corresponds to a value of \l and each row to a value of $V_C$ (in km/s).}
\end{figure*}

An inspection of Fig. 4 shows that this may be a relatively good approximation for
a large part of galaxy's history, but not for the earliest and latest times
(i.e. for the first and last couple of Gyr). We used an exponentially declining SFR
to fit the obtained SFR histories (shown by the  dotted curves in Fig. 4).
The shortest caracteristic timescales correspond to the more massive and compact
disks (larger  $V_C$, smaller $\lambda$): in those cases, the rapid infall makes
the gas available early on, while the large $V_C/R_d$ factor of the adopted SFR leads
to the rapid consumption of that gas. Whether such massive and compact disks exist
is another matter (see  Sec. 4). As we move to larger values of $\lambda$
(i.e. more extended disks with larger $R_d$), the $V_C/R_d$ factor makes
star formation less efficient and the resulting SFR has a smoother history.
In fact, for most of the values of the parameter couple ($V_C,\lambda$) the SFR remains
essentially constant during most of the disk's history, since the gas consumption
is compensated by the infall. Finally, in the case of the most extended and less
massive galaxies (low $V_C$, large $\lambda$) the SFR and gas mass actually increase
in time, due to the adopted prescription for infall (which is very slow for low
mass and low surface density galaxies).  Extended galaxies are found to have 
low surface  brightness (as already shown in Jimenez at al. 1998)
and high gas fractions (see Sec. 3.2).
In summary, it turns out that $\lambda$ {\it affects mainly the shape of star formation
history (i.e. the caracteristic time scale of star formation), while $V_C$ determines
the absolute values of SFR, gaseous and stellar content} (see Sec. 3.4).

The fact that in some cases the SFR increases in time leads us to define the time-scale
$\tau_{\bf *}$,  required for forming the first half of the stars of a given galaxy.
This time scale for star formation appears in Fig. 5, plotted as a
function of $\lambda$ (left) and $V_C$ (right). It can be seen that $\tau_{\bf *}$
is a monotonically increasing function of $\lambda$ and a decreasing function of $V_C$.
For the Milky Way ($V_{CG}$=220 km/s, $\lambda_G$=0.03) we find $\tau_{\bf *}\sim$3.5
Gyr. For galaxies less massive than the Milky Way and $\lambda>0.06$, it  takes more than
half the age of the Universe to      form the first half of their stars.

\begin{figure*}
\psfig{file=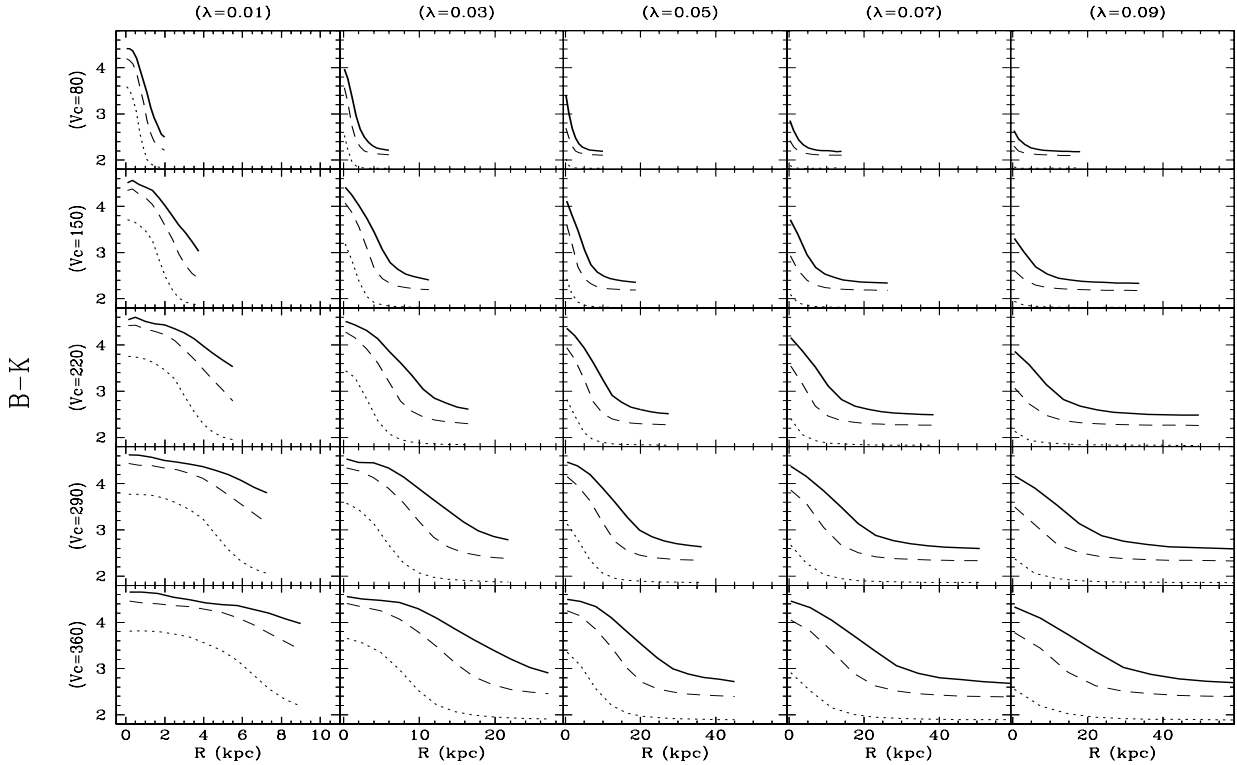,height=11.cm,width=\textwidth,angle=-90}
\caption{\label{profilBK}  Evolution of the B-K profile of our
disk models. Profiles are shown at ages T=3 ({\it dotted}), 7.5 ({\it dashed})
 and 13.5 Gyr ({\it solid}).
Each column corresponds to a value of \l and each row to a value of $V_C$ (in km/s).
Colour profiles generally flatten with time, because of the inside-out scheme
of star formation adopted (see text).}
\end{figure*}

\subsection{Photometric profiles}

In this section we present some results on the photometric evolution of our disk
models. We illustrate this evolution by commenting only on the B-band surface
brightness profile.  Other photometric bands
present similar trends and are not shown here, for lack of space; 
results concerning some of them
will be presented in Sec. 4, where comparison is made to a large body of
observational data.

The evolution of the B-band surface brightness is shown in Fig. 6 for three ages:
t=3, 7.5 and 13.5 Gyr, respectively. It reflects clearly the adopted scheme
of inside-out star formation, i.e. the fact that gas is more rapidly available
in the inner galactic regions (because of the adopted infall timescale
$\tau(R)$), where it is
efficiently turned into stars (because of the $V/R$ dependence of the SFR).
At late times, relatively little gas is left in the inner regions, but quite
a lot is still infalling in the outer ones, maintaining an important SFR there.
Several interesting features should be noticed:

a) The   central surface brightness $\mu_{B0}$ today depends little on $V_C$,
while the more compact (higher $\lambda$) disks have, in general,
higher  $\mu_{B0}$. In almost all cases  $\mu_{B0}$ was higher in the past

b) As a result of the inside-out star formation, the B-band surface brightness
profile flattens, in general, with time. Disks were more compact in the past.

c) Photometric profiles are rarely ``purely exponential''; indeed, they may
deviate considerably from the usually adopted exponential ones. This is in
agreement with the conclusions of Courteau and Rix (1999) who find that
only about 20\% of their sample of more than 500 disk galaxies qualify
as pure exponentials.

d) Because of the inside-out star formation scheme,
 the isophotal radius $R_{25}$ (the radius at which
$\mu_B$=25 mag arsec$^{-2}$) generally increases with time (except for the
$\lambda$=0.01 case). Present day values of $R_{25}$ in our models are found to range
from a few kpc to $\sim$40 kpc.

e) Several of the models in Fig. 6 have $\mu_{B0} <$22.5 mag arcsec$^{-2}$ at
present, i.e. they correspond to Low Surface Brightness (LSB) galaxies. This
concerns all the galaxies with $\lambda$=0.09 and some with $\lambda$=0.07.
However, according to our models,
some of those galaxies have always been LSBs and some of them not. In the
former class belong the less massive galaxies, which were fainter in the past
because of their continuously increasing SFR (Fig. 4). In the latter class
belong more massive galaxies, which have a declining SFR and therefore were
brighter in the past; they became LSBs only after a certain age.
 Finally, in some intermediate cases (e.g. $V_C$=150 km/s and
$\lambda$=0.05, 0.07) the central SFR and corresponding $\mu_{B0}$ have remained
virtually unchanged during the galaxy's lifetime.
These results may have important implications for the study of LSBs as we shall see
in Sec. 3.4.

\subsection {Colour profiles}

Figure 7 presents the evolution of the B-K profiles of our model disks. Again, the
SFR histories of Fig. 4 and the inside-our star formation scheme allow to understand
the main features:

i) The more compact and more massive a galaxy is, the larger is the final B-K
value, either the central   (Fig. 7) or the integrated one (Fig. 8). This is due to
the fact that low $\lambda$ and high $V_C$ values favour a rapid early star formation
and the existence of relatively old stellar populations today.
In fact, the properties of $\lambda$=0.01 galaxies correspond more to those of bulges
or ellipticals than disk galaxies.

ii) B-K values are always larger in the inner galactic regions, since stars are formed
earlier there than in the outer disks. As time goes on,
a ``reddening'' wave propagates outwards in the disks.

iii) More massive disks have a flatter B-K profile than less massive ones
with the same $\lambda$.

iv) The outer galactic regions have, in general, a ``flat'' B-K profile, because they
form stars at late times. Stellar populations in those regions have, on the average,
the same age (they are young) and same B-K values. On the other hand, in the inner
regions there are substantial discrepancies in the timescales of star formation,
and colour gradients are consequently established.

v) For a given $V_C$ the final B-K profile is flatter in more compact disks, since
the $1/R$ factor of the SFR and the infall timescales are such that
efficient early star formation may take place even in their outer regions.

\begin{figure*}
\psfig{file=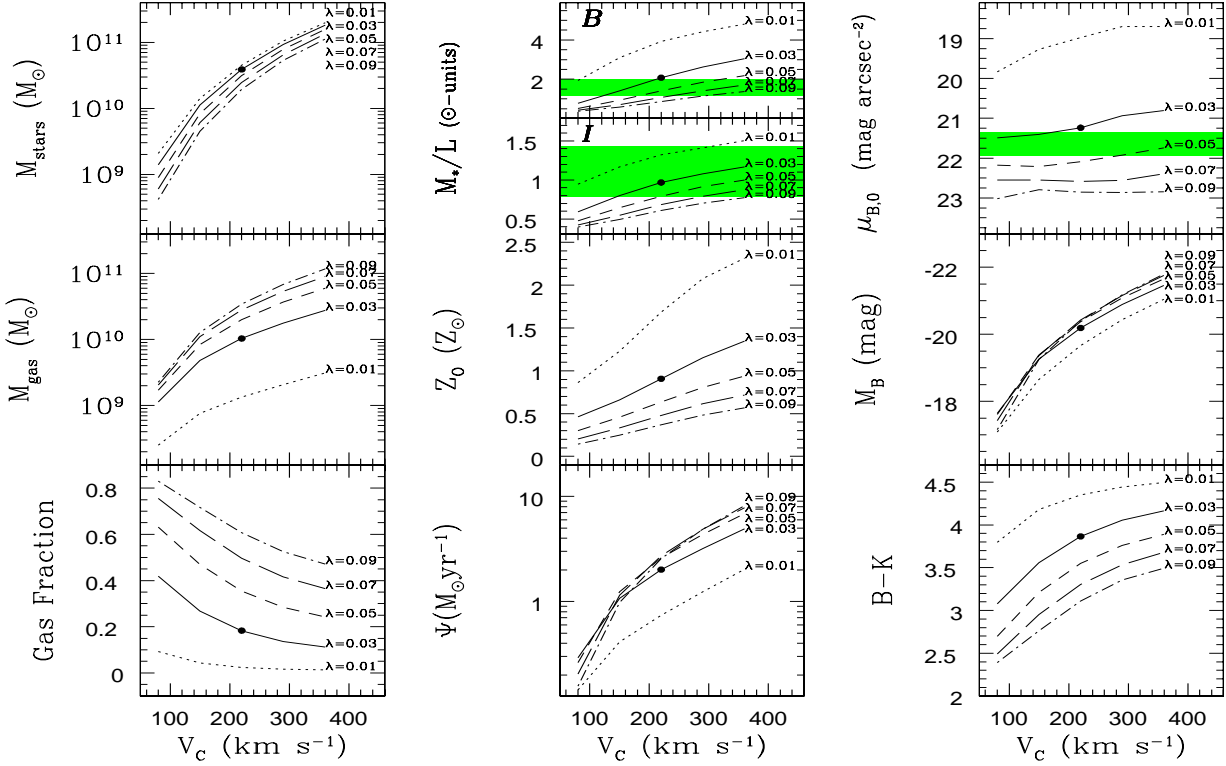,height=11.cm,width=\textwidth,angle=-90}
\caption{\label{FIGint}  Integrated properties of our models at an age T=13.5 Gyr.
Mass of stars ({\it top left}),
mass of gas   ({\it middle left}),
gas  fraction ({\it bottom left}),
M/L values in B- and I- bands ({\it top center}),
average oxygen abundance, i.e. total mass of oxygen divided by total mass of gas
in the disk ({\it middle center}),
star formation rate ({\it bottom center}),
central surface brightness $\mu_{B0}$   ({\it top right}),
total B-magnitude        ({\it middle right}),
total B-K     ({\it bottom right}). All results are plotted as a function of
the rotational velocity $V_C$ and parametrised with the values of the spin
parameter $\lambda$. The {\it filled symbols} on the $\lambda$=0.03 curves
and for $V_C$=220 km/s in all pannels  correspond to Milky Way values; 
as stressed in Sec. 2 the
Milky Way is used to calibrate all our models, through the adopted scaling
relations (eqs. 14 and 15).
The {\it grey bands} in the $M/L$ and $\mu_{B0}$ pannels indicate
the corresponding ranges of observed values (see Sec. 3.4, points 4 and 5).
}
\end{figure*}

The results presented in the previous section and in this one may have
interesting implications for LSB galaxies. Indeed, it can be seen in Figs. 5 and 6 that 
disks with the same $\mu_{B0}$ today may have quite different central colours. This
is e.g. the case for ($\lambda=0.07, V_C$=80 km/s) and ($\lambda=0.07, V_C$=220 km/s).
They both have $\mu_{B0}\sim$23 mag arscec$^{-2}$, but the former has (B-K)$_0$=2.8,
while the latter has (B-K)$_0$=4.  This is due to their different SFR histories,
since in the former case most of the stars are formed late, while in the latter
they are formed quite early on. These results could be related to the conclusions
of a recent study by Bell et al. (1999), who find that there are two kinds of
LSBs, ``red'' and ``blue'', with different star formation histories: the former
are better described by a ``faded  disk'' senario, while the latter are described
by models with low, roughly constant star formation rate. According to our senario,
these different routes are dictated simply by the galaxy's mass, which induces
an earlier SFR in the former case (because of the more rapid infall).
A clear prediction of our models is that ``blue'' LSBs should be less massive and
less luminous than the ``red'' ones.

The observational implications of colour and photometry profiles and their
relationship to abundance gradients (not presented here) is postponed to
a forthcoming paper (Boissier and Prantzos, in preparation).

\subsection{Integrated present day properties}

Assuming  that all our model disks started forming their stars $\sim$13  Gyr ago,
we present in Fig. 8 their integrated properties today (at redshift $z$=0).
Several interesting results can be pointed out:

1) For a given $\lambda$, the stellar and gaseous masses are monotonically
increasing functions of $V_C$. For a given $V_C$ disks may have considerably
different gaseous and stellar contents (by factors 30 and 5, respectively),
depending on how ``compact'' they are, i.e. on their $\lambda$ value.
The more compact galaxies have always a smaller gaseous content.
This is well illustrated by the behaviour of the gas fraction $\sigma_g$({\it bottom left}
of Fig. 8). 
There seems to be a rather good correlation between $\sigma_g$ and $V_C$ for all 
$\lambda$ values, a feature that is indeed observed (see Sec. 4.4).
The most compact galaxies have always $\sigma_g$ lower than 0.1,
while the most extended ones (those corresponding to the LSBs discussed in the
previous section) have always $\sigma_g>$0.5. The latter property seems indeed to
be an observed feature of LSBs (McGaugh and de Blok 1997, and Sec. 4.4).
In general, the final gas fraction is a slowly decreasing function of $V_C$ and
a rapidly increasing function of $\lambda$.

2) The pattern of the average oxygen abundance (i.e. total oxygen mass in the gas,
divided by the gas  mass, {\it central pannel} in Fig. 8) can be readily understood
on the basis of the behaviour of the gas fraction: wherever the latter is low (see
point 1) the metallicity is high. Most of our models have average metallicities
lower than solar. Only the most massive and compact ones  develop average
metallicities larger than 2 times solar. Our LSB candidates have average
metallicities $\sim$0.5 solar.

3) The total SFR presents an interesting behaviour ({\it middle bottom pannel}):
the current SFR $\Psi_0$ is found to be mostly a function
of $V_C$ alone, despite the fact that the amount of 
gas varies by a factor of $\sim$2-3 for
a given $V_C$. This is due to the fact that more extended disks (higher $\lambda$)
are less efficient in forming stars, because of the $1/R_d$ factor (for a given $V_C$).
This smaller efficiency compensates for the larger gas mass available in those galaxies, 
and a unique value is obtained for $\Psi_0$. This immediately implies that little
scatter is to be expected in the B-magnitude vs. $V_C$ relation, and  this is indeed
the case as can be seen in Fig. 8 ({\it middle right pannel}). 
The $\lambda$=0.01 case does not follow this pattern, since very little gas is left
in the disk; as stressed on several occasions, this case corresponds rather
to bulges or elliptical galaxies and is shown here only as a limiting case of our disk
models. The implications
for the Tully-Fisher relation are discussed below (Sec. 4.2).

4) The mass to light ratio $\Upsilon=M/L$ is a quantity often used
to probe the age and/or the IMF of the stellar population and the
amount of dark matter, but also to convert results of dynamical models
(i.e. mass) into observed quantities (i.e. light in various wavelength
bands). From disk dynamics of high surface brightness galaxies
Bottema (1997) inferred $\Upsilon_B$=1.79$\pm$0.48 (for a Hubble 
parameter $h$=0.75). Using the value of $B-I$=1.7 for typical disk galaxies
given by McGaugh and de Blok (1997), MMW98  derived 
$\Upsilon_I$=(1.7$\pm$0.5)$h$, where the Hubble parameter $H_0$ = 100 $h$ km/s/Mpc.

Our model values for  $\Upsilon_B$ and $\Upsilon_I$ are plotted in Fig. 8
({\it middle upper pannel}) along with the corresponding 
observational range ({\it grey bands}).
It can be seen that: a) both $\Upsilon_B$ and $\Upsilon_I$ model values are
slowly decreasing functions of $V_C$ and slowly increasing functions of $\lambda$;
b)  most of the model values lie well inside the observationnaly derived
range; c) the $\lambda$=0.01 case lies outside the observed range of $\Upsilon_B$
(another indication that this value of the spin parameter does not produce 
realistic disks).

In summary, our models lead to fairly acceptable values of 
$\Upsilon_B$ and $\Upsilon_I$,
these values are confined to a relatively narrow range and  a systematic
(albeit weak) trend is obtained with galaxy's mass, that should not be neglected
in detailed models.

5) As discussed in Sec. 3.3, the central surface brightness $\mu_{B0}$ depends little on
$V_C$, but quite strongly on $\lambda$. 
This can be qualitatively understood from Eq. (15)
implying for the central surface density a strong dependence on $\lambda$, but a mild
one on $V_C$. Taking into account that the $\lambda$-distribution peaks around 0.04-0.05 
(MMW98) we see that most of our model disks have a roughly constant current 
$\mu_{B0}\sim$21.5 mag arcsec$^{-2}$. This is quite an encouraging result, in view
of the observed constancy of the central surface brightness in disks, around the
``Freeman value'' $\mu_{B0}$=21.7$\pm$0.3 mag arcsec$^{-2}$ (Freeman 1970); this
value is indicated by the grey band in Fig. 8 ({\it top right pannel}).
In the next section we show that the obtained values of central surface
brightness in longer wavelengths  are also in fair agreement with recent
observational results.
Finally, models with $\lambda$=0.07 and 0.09 result in LSB galaxies today (but not
always in the past, see the discussion in Sec. 3.3).

6) The final B-K colours of our models (Fig. 8, {\it bottom right pannel})
can also be readily understood by an inspection of the gas fraction pattern:
for a given $\lambda$, more massive disks have less gas and are redder, while for a
given $V_C$ more extended disks (larger $\lambda$) have more gas and are bluer.
Overall, the less massive galaxies have B-K values in the 2.5-3 range, while for
the most massive ones we obtain values around 3.5-4.

\section{Results vs observations}

Having described in the previous section the main results of our models in terms
of the input parameters $V_C$ and $\lambda$, we turn now to a comparison of these
results to observations. Our purpose is to check whether  the results of our grid
of models (a) bracket reasonably well the range of observed values of various
galactic properties and  (b) reproduce some observed strong correlations.
Such a comparison is, in principle, possible if the age of the observed galaxies
is known, but this is unfortunately not the case. We shall make then the
assumption that all galaxies started forming their stars at the same epoch,
around 13 Gyr ago. It is not {\it a priori} obvious
whether this assumption is reasonable or not, but we shall see that in most
cases it leads to quite acceptable results. In any case, as stressed in
Sec. 3.1 the term ``time of formation'' is a rather loose concept for models
with infall: the ``effective'' timescales for the formation of the stellar
populations in our models are given in Fig. 5.

\subsection{Boundary conditions: 
Disk size  and  central surface brightness}

As explained in Sec. 2.2, our model disks are described by the central surface density
$\Sigma_0$ and scalelength $R_d$. These are related to the ``fundamental''
parameters $V_C$ and $\lambda$ (fundamental because they refer to the dark
matter haloes) through the scaling relations (14) and (15), which allow
to use the Milky Way as a ``calibrator''.

In this section we check whether this type of modelling produces present
day disks with ``reasonable'' morphological properties. The central surface brightness
and scalelength in the r-band represent fairly well the underlying stellar
population and are less affected by uncertainties due to dust extinction.
The behaviour of our models in shorter wavelengths is somehat different, 
but we postpone an analysis of the colour gradients of our models to a
forthcoming paper.

\begin{figure}
\psfig{file=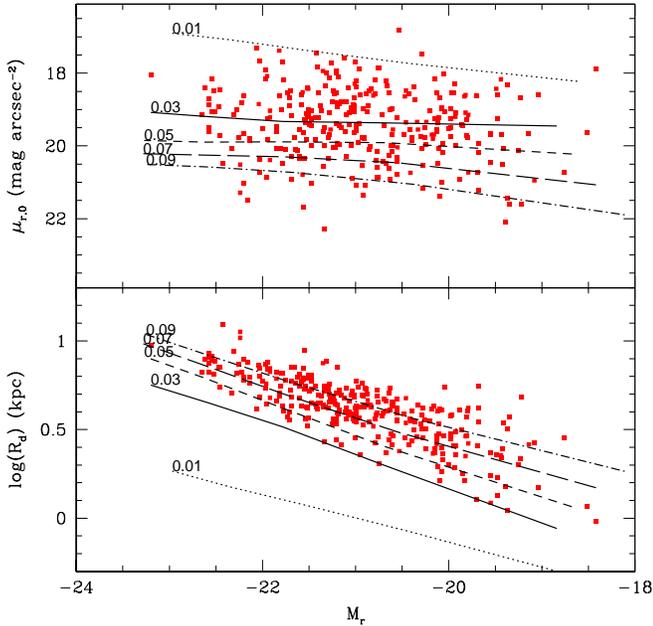,height=9.cm,width=0.5\textwidth}
\caption{\label{CSBSIZE}  Central surface brightness ({\it upper pannel})
and disk size $R_d$ ({\it lower pannel}) in the r-band, as a function
of the galaxy's magnitude $M_r$. Our results are parametrised by the
values of the spin parameter $\lambda$; $\lambda$=0.01 models produce
unrealistically small  disks with very high surface brightness and correspond rather
to galactic bulges. Data are from Courteau and Rix (1999).
}
\end{figure}

In a recent paper, Courteau and Rix (1999) analyse two large data samples
of more than 500 non-barred disk galaxies. They find that only $\sim$20\%
of these disks can be described by purely exponential profiles and they
provide disk sizes, rotational velocities, magnitudes and colours.
Their data allow to derive the central disk surface brightness.

In Fig. 9 we present the r-band central surface brightness (upper pannel)
and size (lower pannel) of our model disks as a function of the galaxy's
luminosity and we compare them to the data of Courteau and Rix (1999).
It can be seen that:

i) The central surface brightness in our models compares fairly well to
the data. For a given value of $\lambda$, $\mu_{r0}$ is quasi-independent on
luminosity (or $V_C$), as explained in Sec. 3.4. The obtained values depend
on $\lambda$, the more compact disks being brighter. All the observed
values are lower than our results of $\lambda$=0.01 models. Higher $\lambda$ values
bracket reasonably well the data.

ii) The size of our model disks decreases  with decreasing luminosity, with
a slope which matches again the observations fairly well. Our disk sizes range 
from $R_d\sim$6-10 kpc for the most luminous galaxies to 1-2 kpc for the less
luminous ones. The $\lambda$=0.01 curve clearly lies too low w.r.t. all data
points, showing once more that it does not produce realistic disks.

It should be noticed that a similar comparison between theory and observations
is made in MMW98. However, for lack of an evolutionary model, MMW98 compare
rather the profiles of their gaseous disks (i.e. those given by our 
eq. 12 and 13) to the data, making the implicit assumption that gas is
transformed into stars with the same efficiency in all parts of all disks.
Our models show that this is a good approximation for the inner regions, but
not for the outer ones (because of the adopted radial dependence of the SFR 
efficiency) and that the final stellar disk sizes are shorter than 
the initial gaseous ones. Moreover, MMW98 adopt a $M/L$ ratio (the same for
all disks) in order to convert their mass surface density to surface 
brightness. As discussed in point 4 of Sec. 3.4, the $\Upsilon=M/L$ ratio is
a slowly increasing function of a galaxy's mass (at least accordingly 
to our models).
Our self-consistent chemical and spectrophotometric evolution
models allow to avoid these approximations and to compare directly the output
to observations.

\subsection{Dynamics: the Tully-Fisher relation}

The Tully-Fisher (TF) relation is a strong correlation exhibited
by the whole population of disk galaxies. It relates their luminosity  $L$
(or magnitude M) and their circular velocity $V_C$ (or $H_{\alpha}$ line width $W$)
and can be expressed as
M$=-a \, [log(2 V_c)-2.5] + b$ or $L \ \propto \ V_c^{a/2.5}$.
The exact slope of the relation and its zero point are still the subject of considerable
debate (e.g. Giovanelli et al. 1997). 
For this reason, we  compare our results with 
several recent fits to the I-band Tully-Fischer relation
obtained with various data samples.

In a recent study, Giovanelli et al. (1997) compiled a homogeneous set 
of I-band photometry and HI velocity profiles of 555 spiral galaxies
in 24 clusters with redshift extending to $cz\sim$9000 km/s. 
The slope of their TF relation is 7.68 $\pm$ 0.13.
They estimate the mean amplitude of the scatter in the observed relation to be
0.35 mag, but they notice that it increases with decreasing velocity. They also
estimate that the combination of statistical and systematic uncertainties can
affect the zero point of the relation up to 0.07 mag, i.e.  a rather small effect.
In order to compare our results to the Giovanelli et al (1997) sample we
shall assume that $h=0.65$ i.e. a value very close to the $h=0.69\pm0.05$
value derived by these authors on the basis of their data.

\begin{table}
\begin{tabular}{l  c c  c }
\hline
        & $a$ (slope) & $b$ (0-point)  & $\sigma$ (scatter) \\
\hline
{\it Observations    }  &            &               &           \\
Giovanelli et al.		& 7.68       & -21.93        & 0.35 mag  \\ % a +- 0.13
Tully et al.			& 8.17       & -21.54        & 0.42 mag  \\	
Han \& Mould		& 7.87       & -21.42        & 0.40 mag  \\ %0.16
Mathewson		& 6.80       & -21.72        & 0.43 mag  \\ %0.08
{\it Our models      }  &            &               &           \\
$\lambda$=0.01          & 6.81/7.17   & -21.34/-21.70 &           \\
$\lambda$=0.03          & 7.01/7.31   & -21.47/-21.72 &           \\
$\lambda$=0.05          & 7.37/7.71   & -21.35/-21.54 &           \\
$\lambda$=0.07          & 7.79/8.10   & -21.16/-21.30 &           \\
$\lambda$=0.09          & 8.18/8.44   & -21.00/-21.05 &           \\
\hline
\end{tabular}
\caption{\label{TABfit}The Tully-Fisher relation in the I-band: observations
vs. our models
(see Sec. 4.2 for references). A value of $h$=0.65 is assumed for the Hubble parameter
in this Table and Fig. 10. Two values (separated by /) are given for our models: 
the first one corresponds
to models with extinction taken into account and the second to models with extinction neglected.
Observational data are from 
Giovanelli et al. (1997), 
Tully et al. (1998), whereas the Han and Mould and Mathewson et al. data are
taken from the compilation of Willick et al. (1996).
}
\end{table}

\begin{figure}
\psfig{file=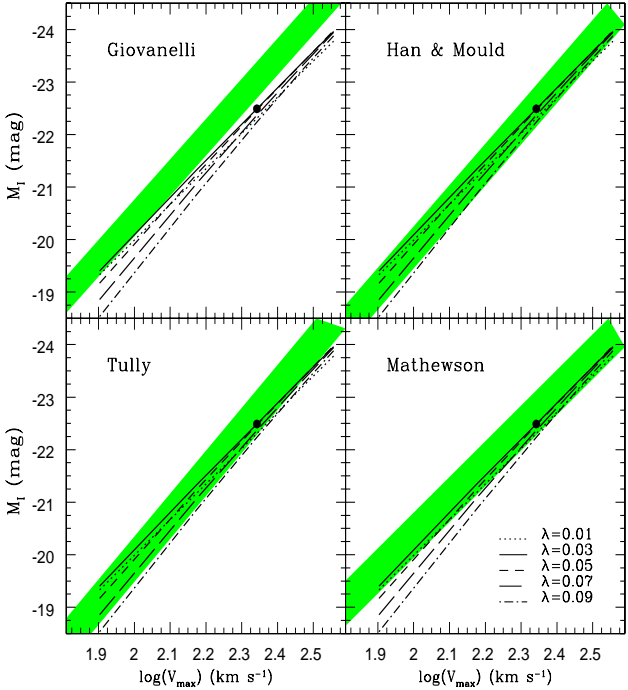,height=10.cm,width=0.5\textwidth}
\caption{\label{FIGTUL} The Tully-Fisher relationship in the I-band. In
each pannel, our results ({\it curves} parametrised by the corresponding $\lambda$
value, as indicated in the {\it bottom right} pannel) are compared with fits from various
authors ({\it shaded zones}), differing by their zero-points and slopes 
(see Table 1 and  Sec. 4.2 for references on the data).
The models shown here are those with extinction taken into account
(first column in Table 1), which lead to flatter TF relations than when extinction
is neglected. 
In all cases (except for Tully et al. 1998) the data are plotted 
assuming that the Hubble parameter $h$=0.65.
}
\end{figure}

Willick et al.\shortcite{willick96} have computed several
TF relations on the basis of samples from various authors. Two of those
samples are in the I-band and shall be used in the present work.
The first one is a cluster sample from Han, Mould and coworkers
(references are given in Willick et al.\shortcite{willick96}). 
They find a slope $a=7.87 \pm$ 0.16 and a scatter of 0.4 mag.
The second one is the field galaxies sample of Mathewson et
al. \shortcite{mathewson92}, composed of  1219 galaxies. It  provides
a slope  $a=6.80 \pm$ 0.08 and a scatter of 0.43 mag.
For these two samples Willick et al.\shortcite{willick96}
define the ``Hubble absolute 
magnitude'' of a galaxy,  computed from its apparent magnitude
and redshift and independent of the Hubble parameter;
the zero-points of the TF relation are then determined through  this 
Hubble absolute magnitude.
In order to plot those two relations on a common scale with other
samples, we transformed the quoted ``Hubble absolute magnitudes'' into usual
absolute magnitudes by assuming $h$=0.65.

A fourth work (Tully et al.\shortcite{tully98}) uses a  limited
sample of 87 galaxies  from the Ursa Major and Pisces Clusters. 
The obtained slope is steeper than in the other
determinations ($a$=8.17) while the scatter is similar (0.42 mag).
In that work, TF relations in other  photometric bands 
(B,R,I,K')  are considered for  
the same galaxy sample; we shall compare them to our results in the end of this section.

The parameters of those various fits are presented  in 
Table \ref{TABfit}, and the data are compared with our
models in Fig. 10. It can be seen that:

a) The model slopes and zero-points depend somewhat on $\lambda$: in general,
larger $\lambda$ values lead to steeper M-$V_C$ relations. This is
is due to the following reason: At high $V_C$ all galaxies transform most of
their gas into stars, i.e. the final stellar mass depends little on $\lambda$
and so is the case for the corresponding M$_I$. As we go to lower $V_C$,
 the final stellar mass
depends more and more on $\lambda$, because higher  $\lambda$ galaxies are more 
inefficient in forming stars: this is reflected in  a steeper slope of the
corresponding M-$V_C$ relation.

b) Since the most probable values of $\lambda$ are in the 0.04-0.05 range
(according to numerical simulations), our corresponding slopes  of the
M-$V_C$ relation are $a\sim$7.1-7.3 (with extinction taken into account). 
This value of $a$ is intermediate between
those obtained from the samples of
Giovanelli et al., Mathewson et al. and Han, Mould and collaborators.
It is certainly 
lower than the slope in the Tully et al. sample, which 
concerns, however, disk galaxies in clusters; in such environments disks
probably evolve quite differently than in the field, being affected by
interactions. Our models are better compared to field galaxies, such as those
in the sample  of Mathewson et al.

c) Because of the very inefficient star formation at low $V_C$ and high
$\lambda$, the TF relation of our models shows a magnitude dependent scatter,
from $\sim$0.1 mag at high $V_C$ to $\sim$0.5 mag at low $V_C$. This trend
is not apparent in the observed relations appearing in Fig. 10
({\it shadowed zones}), since only the
average scatter appears.  However, Giovanelli et al. (1997) find that there is 
indeed a trend of increasing scatter with decreasing magnitude in the TF 
relation. Notice that in our models, the galaxies that contribute mostly
to that scatter at low $V_C$
are the low surface brightness galaxies ($\lambda \sim$0.07-0.09); such 
galaxies are, in general, not included in the observed samples. In any case, 
a clear prediction of our models is that the scatter in the TF relation should
be more important at low $V_C$.

d) Extinction modifies somewhat the slopes, leading to slightly swallower 
TF relations than those obtained by the stellar population alone.
This is because it affects more the  massive (high surface brightness and
high metallicity) disks, rather than the low mass ones.
If our extinction values are overestimated, then the slopes of our models
should be intermediate between the two values given in Table 1. This is also
true for the zero-points.

It should be noticed here that, if the initial gas were almost completely turned 
into stars, one should naturally expect a $L \ \propto \ V_C^3$ relation
(from Eq. 4, assuming $m_d =const$ and $M/L =const$). In other words, the adopted
scaling relations lead to a ``built-in'' TF relation, as properly pointed out
in MMW98. However, when star formation is properly taken into account in 
``realistic models'', different slopes are found, depending on the SFR efficiency
(itself a function of $\lambda$, see Table 1). Although   the differences introduced
by the more realistic senarios are relatively small, they may have important 
implications for a proper interpretation of the TF relation. In particular, our
models, taken at face value, suggest that low surface brightnesss galaxies (i.e. those
with $\lambda>$0.07) should have steeper TF relations than those of high surface
brightness in the I-band.

\begin{figure}
\psfig{file=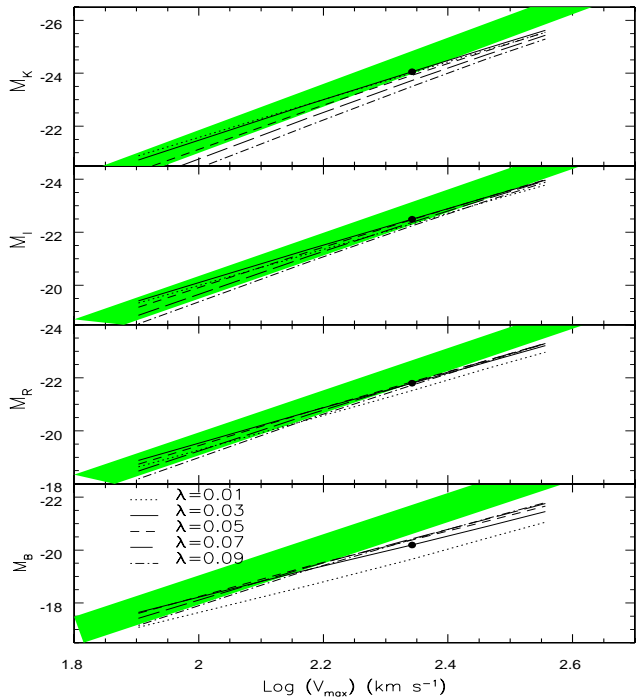,height=10.cm,width=0.5\textwidth}
\caption{\label{FIGTUL2} The Tully-Fisher relationship in several photometric bands. 
Our results  ({\it curves} parametrised by the corresponding $\lambda$
value, as indicated in the {\it bottom} pannel)
are compared with the Tully et al.\shortcite{tully98} observations
(shaded zones).}
\end{figure}

% >>> JE NE SUIS PAS SUR QUE BONNE IDEE >> VALEUR OBSERVEE...
\begin{table}
\begin{tabular}{ l l l l l l l }
\hline
Photometric & \l=0.01      & 0.03      &     0.05 &     0.07  &   0.09      & Observed \\
band        &              &              &             &              &    & (Tully)   \\
\hline
B                 &    6.1      &  5.8        &   6.2       &   6.6       &    7.1 & \emph{7.79} \\
R                 &    6.7      &  6.7        &   7.0       &   7.4       &    7.9 & \emph{7.96} \\ 
I                 &    6.8      &  6.9        &   7.3       &   7.8       &    8.2 & \emph{8.17} \\
K'                &    7.1      &  7.5        &   8.0       &   8.4       &    8.8 & \emph{8.73} \\

%[Ces donnees > _infallD ]

\hline
\end{tabular}
\caption{\label{TABLETUL}Slopes in the TF relations $L_{\lambda} \propto  V_c^{a/2.5}$, in 
various wavelength bands. 
Our models (arranged by $\lambda$ values) are compared to the observations of Tully et al. (1998);
see also Fig. 11.}
\end{table}

Tully et al. (1998) also investigated the slope of the TF relation in several 
wavelength bands. They found that the slope increases with wavelength, as can be seen
in Table 2. Our results are also shown on Table 2 and in Fig. 11, compared to the
observational fits of Tully et al. (1998). The following points should be noticed:

i) In our models, the trend of increasing slope with increasing $\lambda$ 
is found in all wavelength bands, not only in the I-band.

ii) As we go from the B- to the K'-band, our models do predict a steeper and steeper slope
for all $\lambda$ values. Leaving aside the $\lambda$=0.01 case (unrealistic for
disks) one sees that the slope $a$ increases between B- and K'-bands 
by $\sim$1.8 for all $\lambda$ values.
This increase is somewhat larger than the corersponding increase of $\sim$1 in the data
of Tully et al. (1998).

iii) Our model values for the slope $a$ are, in general, smaller than those of Tully et al.
(1998), especially in the B-band. Only the $\lambda$=0.09 models lead to values comparable
to those of Tully et al. (1998) in the R, I and K' bands. 

As stressed already, the Tully et al. (1998) data concern spirals in clusters which cannot
be directly compared to those modelled in our work; our slopes in the I-band compare
rather well to the field galaxy sample  of Mathewson et al. But the important issue here
is the trend of {\it increasing slope} with wavelength, found in both theory and observations.
In our models this increase is due to the fact that less massive galaxies are chemically
and photometrically ``younger'' than their more massive counterparts, having in general
``bluer'' colours. This is shown in Fig. 8 ({\it bottom right} pannel), where B-K
increases systematically with $V_C$. This change in B-K with $V_C$ produces a steepening
in the slope of the TF relation as we go from the B- to the K-band. This feature is not
``built-in'' in our models (it does not stem directly from the adopted scaling relations) but
it results from the adopted prescription for the SFR {\it and} the infall rate (Sec. 2.3),
making less massive galaxies appear ``younger'' than more massive ones. This 	
generic feature of our models is crucial also in interpreting the metallicity-luminosity
relation of disks (see Sec. 4.5) and we consider it to be one of the most important
points of this work.

In a recent paper, Heavens and Jimenez (1999) use also semi-analytic disk models
in CDM senarios and a rather simple prescription for star formation in order to 
investigate the
variation of the slope of the TF relation with wavelength. They apparently manage to 
reproduce the oberved trend, but they offer no explanation  for their success.
They also suggest that $\lambda$ value and formation redshift $z$ are the parameters that
can explain the observed dispersion in the TF relation. Our models show that this
is certainly true for $\lambda$, but the disk baryonic fraction $m_d$ 
(taken as constant here) could obviously play a similar role.

\begin{figure*}
\psfig{file=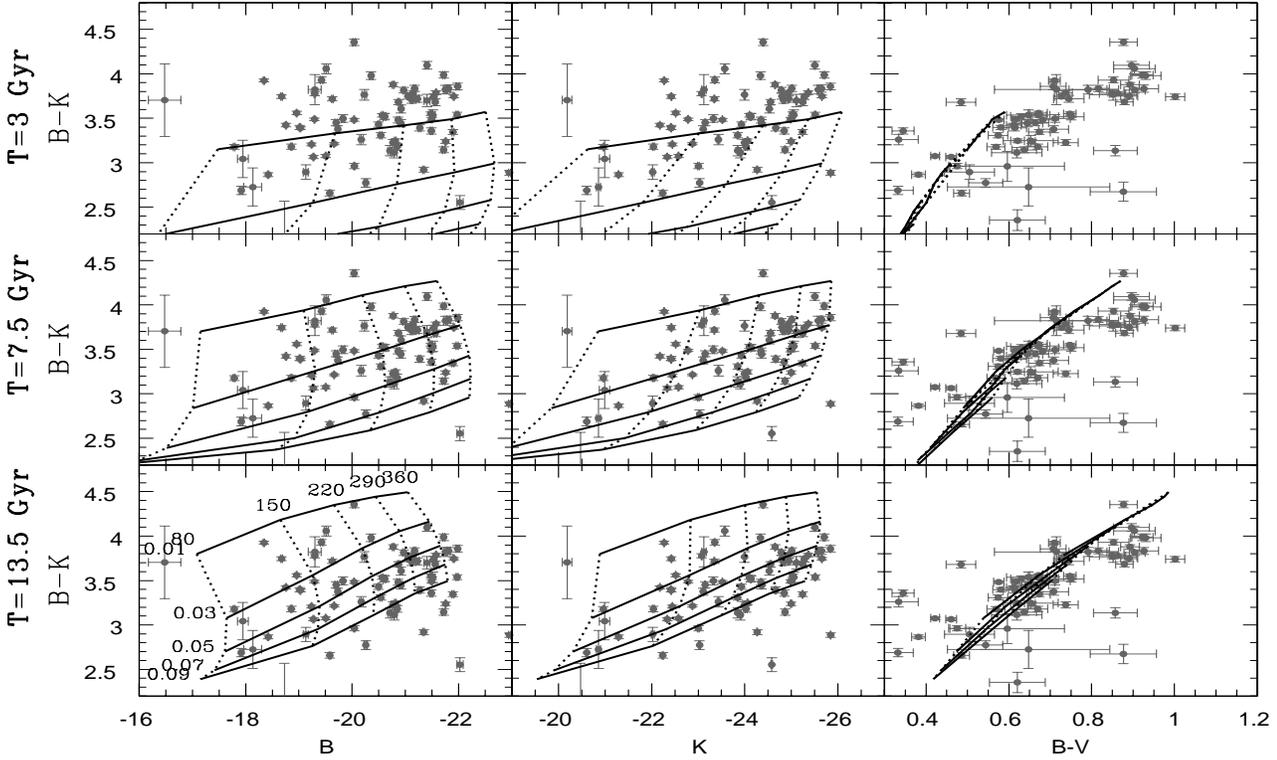,height=11.cm,width=\textwidth,angle=-90}
\caption{\label{COLMAG} 
Colour-magnitude and colour-colour relations for our models
at three different ages:  3 Gyr ({\it upper pannels}), 7.5 Gyr
({\it middle pannels}) and 13.5 Gyr ({\it lower pannels}).
The grid of models represents 5 values for the spin parameter $\lambda$
(0.01 to 0.09 from {\it top to bottom} in {\it solid curves}) 
and 5 values for the circular velocity  $V_C$
(80 to 360 km/s from {\it left to right}, in {\it dashed curves}), 
arranged as indicated in the lower left pannel.
In the $B-K$ vs $B-V$ relation our grid is ``squeezed'' onto a relatively
tight relation, determined mainly by $V_C$.
Data in all pannels (the same points are plotted in each row) 
are from de Jong (1996).
We have calculated magnitudes and colours for {\it disks only}, using
de Jong's fits (central surface brightness and disk scalelength). 
Clearly, the observed galaxies
are older than 3 Gyr, and rather closer to 13.5 than to 7.5 Gyr.}
\end{figure*}

\subsection{Luminosity evolution: 
Colour-colour and colour-magnitude relations}

Spectro-photometric models are usually calibrated on colour-colour and
colour-magnitude diagrams. The parameters of the models (star formation
efficiency, stellar IMF, infall time-scales, ages ...) are chosen as to reproduce
the properties of galaxies of all morphological types
(e.g. Lindner et al.\shortcite{lindner99}, % Ou bien Moller 97 AetA 317 676
Fioc \& Rocca-Volmerange\shortcite{fioc97}). 

As explained in the Introduction, our study is limited to the case
of spiral galaxies, because they offer a much larger amount of
observational data to compare with. However,
we have not ``tuned'' any parameters in order to reproduce
any diagram of this kind. It is thus of interest to see if our models,
generalized from the Milky Way by the use of simple scaling relations,
are able to reproduce those
diagrams for spiral galaxies. For this aim, we used data from De Jong, 1996.
De Jong's data consists in radial and integrated photometry
of 86 undisturbed face-on spirals. 
They are thus directly comparable to our models, since  uncertainties
due to inclination, or to perturbations by nearby galaxies
(which could have affected the star formation activity) are avoided.
We evaluated the disk magnitudes and colours in de Jong's sample by using his fits 
for the disks only (i.e. central surface brightness and disk scalelength) 
and plot them on Fig. 12 (notice that the error bars reflect the uncertainties
of the fit, not those of the observations). 

Since the age of the observed galaxies is not known, we show in Fig. 12
our results for three different epochs: t=3, 7.5 and 13.5 Gyr, covering
the expected range of galaxy ages. It is clearly seen that:

1) At all ages there is a tight $B-K$ vs $B-V$ correlation, 
quasi-independent of the value of $\lambda$. It depends only on the
galaxy's mass (i.e. luminosity): lower mass galaxies are always
``bluer'' than more massive ones.

2) Results at t=3 Gyr do not compare well to the data: model disks are
too blue. The observed galaxies cannot be as  be as young as 3 Gyr.

3) At t=7.5 Gyr the results compare much better to the data, although there
is still a lack of ``red'' model galaxies 
(i.e. with $B-V$ in the 0.8-1.1 range).

4) At t=13.5 Gyr the agreement between theory and observations is quite
 satisfactory. The observed colours and magnitudes are well bracketed by 
our grid  of results. 

5) Our models do predict a colour-magnitude relation which becomes steeper with time.
More luminous galaxies are ``redder'' on average (as can also be seen on
Fig. 8, {\it bottom right}). De Jong's data support this picture qualitatively.
Also, the slope of the B-K vs. B-V relation ({\it bottom right pannel} in Fig. 12)
depends little on $\lambda$ or $V_C$; it is slightly steeper than the
observed one, but the overall comparison is fairly satisfactory.

It is obviously difficult to assign an ``age'' to these galaxies on the
basis of such diagrams. Increasing slightly the efficiency of star formation
in Eq. (1) would lead to similar results at ages lower by a few Gyr.
Further constraints should be considered, like those analysed in the next sections.

We wish to emphasize here the fact that our adopted senario (infall and star formation
timescales longer for low mass galaxies) leads naturally to the trend of 
massive galaxies being ``redder'' on average, as observed. This trend is a well-known
problem in hierarchical models of galaxy formation, where low mass galaxies are formed
first and are found to be ``redder'' today than massive ones.
In their recent work Somerville and Primack (1998) suggest that this problem may
be overcome by including in the models dust extinction, which affects more the
massive galaxies and makes them ``redder'' on average. However, in our models we find
that dust extinction plays a minor role: it can modify slightly the situation (enhancing
the already existing trend in the stellar population alone) but not reverse it
completely. And is certainly not dust extinction that makes the Milky Way appear
chemically and photometrically older than the Magellanic Clouds.

\begin{figure*}
\psfig{file=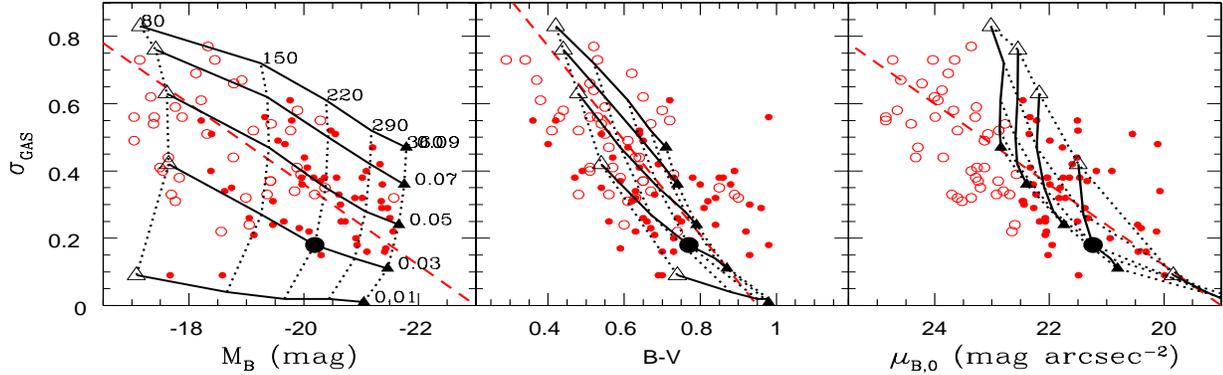,height=11.cm,width=\textwidth,angle=-90}
\vspace*{-5.5cm}
\caption{\label{gascol}  Gas fraction vs. B-magnitude ({\it left pannel}),
B-V colour ({\it middle pannel}) and central surface brightness
$\mu_{B0}$ ({\it right pannel}).
The grid of models represents 5 values for the spin parameter $\lambda$
({\it solid curves}) and 5 values for the circular velocity
({\it dashed curves}), as indicated on the left pannel.
Values of $V_C$ run from 80 km/s ({\it open triangle}) to 360 km/s
({\it filled triangle}).
Results of models are given at a galactic age of 13.5 Gyr.
Data in all pannels are from McGaugh and DeBlok (1997).
{\it Filled symbols} correspond to high surface brightness galaxies and
{\it open symbols} to low surface brightness galaxies.
Our grid of models clearly does not fit the lowest surface brightness
galaxies (i.e. below $\mu_{BO}$=23 mag arcsec$^{-2}$), which probably
require values of $\lambda >$0.1.
In all three pannels, the {\it dashed} diagonal line is a fit to the
data given in McGaugh and DeBlok (1997): $\sigma_g$ = 0.12 (M$_B$+23) =
-1.4 [(B-V)-0.95] = 0.12 ($\mu_{BO}$-19).
}
\end{figure*}

\subsection{Chemical Evolution: 
Gas fraction vs. Magnitude and Colour}

Diagnostics of galaxy evolution can be made in a variety of ways. Colour-colour
diagrams are often used for that purpose, as discussed in the previous section.
However, their usefulness is limited by the fact that they admit 
``degenerate solutions'', i.e. ``red'' stellar populations may result either
from old age and low metallicity or young age and high metallicity.

A key ingredient to galaxy evolution is the consumption of gas, turned into 
stars; as time goes on, the gas fraction $\sigma_g$  decreases
in general (except in cases where infall exceeds the star formation rate).
The gas fraction could then, in principle, be used to probe the evolutionary
status of galaxies.

This probe has rarely been used in studies of galaxy evolution up to now,
for several reasons. First, it is also affected by ``degeneracy'': galaxies
may have low gas fractions either because of quiescent star formation for
a long time  or because of recent enhanced star forming activity. Secondly,
until recently, the  observed range of values in $\sigma_g$ was rather small
($\sigma_g <$0.2) and provided little constraints to models. It seemed
indeed that we live at a special epoch, when most galactic disks have almost
exhausted their reservoir of gas (e.g. Kennicutt et al. 1994).

In an important recent work McGaugh and deBlok (1997) compiled data from 
several surveys and derived various important quantities for galaxy evolution
studies: colours, absolute magnitudes, scalelengths, central surface
brightness, atomic and molecular gas amounts. On the basis of these data
they were able to establish several important correlations between the
gas fraction and other quantities, in particular B-V and B-I colours,
absolute magnitudes and central surface brightness in the B- and I- band.
On the other hand, they found no apparent correlation between gas fraction and
disk scalelength. Another important finding of their work is that gas rich
galaxies (i.e. with $\sigma_g >$0.4) do exist, but they are systematically
low surface brightness galaxies.

 In Fig. 13 we compare our results to the data of McGaugh and deBlok (1997).
The following points can be made:

1) We find indeed a trend between gas fraction and absolute magnitude: smaller
and less luminous galaxies are in general more gas rich, as can also be seen
in Fig. 8 ({\it bottom left} pannel). Our grid of models covers reasonnably 
well the range of observed values, except for the lowest luminosity galaxies 
(see also points 2 and 3 below). In particular, the slope of the
$\sigma_g$ vs. M$_B$ relation of our models is similar to the one in the fit
to the data given by McGaugh and deBlok (1997). Moreover, among our models, 
closest to this fit are those with $\lambda$=0.05 (and 0.04, if we
interpolate to the next value of $\lambda$=0.03), i.e. the value near the peak
of the $\lambda$-probability distribution that is obtained from numerical
simulations.

2) We find that the observed $\sigma_g$ vs. B-V relation can also be
explained in terms of galaxy's mass, i.e. more massive galaxies are
``redder'' and have small gas fractions. Indeed, 
the slope of the $\sigma_g$ vs. B-V relation (-1.4) is again well reproduced
by our models, and the $\lambda$=0.05 models lead to results that match close
the observations. However, although our grid of models reproduces well the 
observed dispersion in $\sigma_g$, it does not cover the full range of B-V
values. The lowest B-V values correspond to gas rich galaxies ($\sigma_g\sim$
0.5-0.8) and could be explained by models with higher $\lambda$ values than
those studies here (i.e. by very Low Surface Brightness galaxies). 
However, observed galaxies that are both ``red''
(B-V$\sim$1) and gas rich ($\sigma_g >$0.3) are difficult to explain in the
framework of our models.

3) The situation is radically different for the $\sigma_g$ vs. $\mu_{B0}$ 
relation. As already seen in Fig. 8 ({\it top right} pannel) we find no
correlation in our models between  $\mu_{B0}$ and $V_C$, but we do find
one between $\mu_{B0}$ and $\lambda$: more compact disks (lower $\lambda$)
have higher central surface brightness. It is this latter property that allows 
our models to reproduce, at least partially, the observed $\sigma_g$ vs. 
$\mu_{B0}$ relation ({\it right pannel} of Fig. 13). It can be seen that for
a given $\lambda$, similar values of $\mu_{B0}$ are obtained for all galaxy
masses (or $V_C$), but  when different values of $\lambda$ are considered, a
large part of the observations in the $\sigma_g$-$\mu_{B0}$ plane
is covered by our grid of models.
Galaxies with high $\mu_{B0}$ (20-21 mag arcsec$^{-2}$) and low $\sigma_g$
correspond to models with $\lambda\sim$0.02. As in point 2 above, however,
we are unable to reproduce galaxies with both high $\mu_{B0}$ and $\sigma_g >$
0.3. On the other hand, our models do not reproduce very low surface brightness
disks, with $\mu_{B0}>$23 mag arcsec$^{-2}$; 
as explained in the previous paragraph, this
is a rather minor problem, since such disks could result from models with
$\lambda>$0.1, not considered here.

We should like to stress that it is the first time (to our knowledge) that
a comparison of this kind of data is made to fully self-consistent models
of chemical and spectrophotometric evolution of galaxies. Although the
agreement between theory and observations  is by no means perfect, it 
nevertheless suggests that the basic picture may be correct. The main result
is that the $\sigma_g$ vs. M$_B$ and $\sigma_g$ vs. B-V relations can
be explained in terms of galactic mass (or, equivalently, $V_C$), but the
$\sigma_g$ vs $\mu_{B0}$ relation can only be (partially) explained in 
terms of the spin parameter $\lambda$.

It is interesting to notice that in their work McGaugh and deBlok (1997)
have speculated on these relations, on the basis of older, one-zone models
of galaxy evolution (by  Guiderdoni and Rocca-Volmerange 1987). In particular,
they anticipated  that the galaxy's baryonic mass may be the driving factor
in explaining the $\sigma_g$ vs. M$_B$ relation, as we find in our models.
However, they suggested that the $\sigma_g$ vs. 
$\mu_{B0}$ relation could be explained if low surface brightness galaxies are
on the average 3-5 Gyr younger than the high surface brightness ones.
Instead, we interpret the observed correlation in terms of $\lambda$, not of
age differences. We are aware though that age is a third parameter that
affects the results and may render the  picture more complex than scetched 
above. We prefer, however, to keep a minimum of simplicity in our models
and assume that all disks started forming stars at least 10 Gyr ago,
unless there is compelling evidence for the contrary.

\begin{figure}
\psfig{file=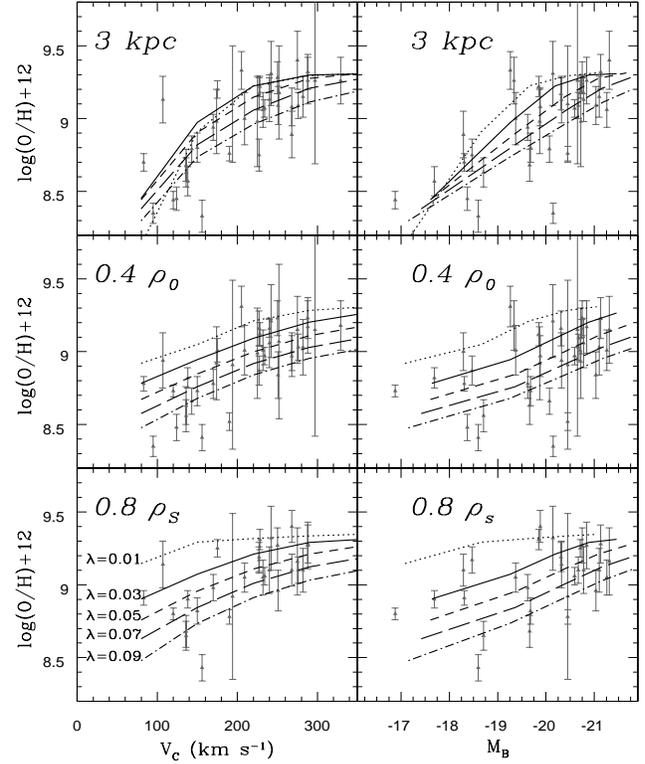,height=11.cm,width=0.5\textwidth}
\caption{\label{ZAR}  
Oxygen abundances at various galactocentric distances:
at 3 kpc ({\it upper pannels}), 0.4 $\rho_o$ ({\it middle pannles}),
and 0.8 $\rho_S$ ({\it lower pannels}). They are plotted as a function
of $V_C$ ({\it left}) or, equivalently $M_B$ ({\it right}).
Our results ({\it curves}) are parametrised by the corresponding
 $\lambda$ values. Data are from Zaritsky et al. (1994).
}
\end{figure}

\subsection{ Chemical evolution: 
Metallicity vs. integrated and local properties}

Metallicity is another probe of the evolutionary status of a galactic disk.
Both its absolute value and its radial profile can be used as a diagnostic
tool to distinguish between different models of galactic evolution.

Zaritsky et al. (1994) provided a sample of 39 disk galaxies for which 
oxygen abundances have been measured in at least 5 HII regions. They determined
radial abundance profiles and deduced the oxygen abundance at three
caracteristic radii: an absolute physical radius (at 3 kpc from the center)
and two dimensionless radii, namely 0.4 $\rho_0$ ($\rho_0$ being the isophotal
radius) and 0.8 $\rho_s$ ($\rho_s$ being the disk exponential 
scale-length, i.e. $R_d$ in our notation). 
Zaritsky et al. (1994) found a strong correlation between the abundances at 
each of these radii and the disk circular velocity $V_C$, and a slightly weaker
correlation between abundances and absolute magnitude M$_B$. They also found a
correlation between abundances and Hubble type, but the fact that mass (i.e. 
$V_C$) and Hubble type are correlated with each other among spirals makes
difficult to evaluate the role of the latter in shaping these correlations.

In Fig. 14 we plot our results at the same caracteristic radii as Zaritsky
et al. (at 3 kpc, at 0.4 $\rho_0$ and at 0.8 $R_d$, from {\it top} to
{\it bottom}, respectively) as a function of circular velocity $V_C$ ({\it
left pannels}) and of M$_B$ ({\it right pannels}).

It can be seen that a correlation is always found in our models between
abundances ($Z$) and  $V_C$ or M$_B$, stronger in the case of $Z$(3 kpc) than
in the other two cases. Only in the $\lambda$=0.01 models (unrealistic for 
disks, as stressed already), $Z$(0.8 $R_d$) is not correlated with either $M_B$
or $V_C$. Except for that, the grid of our models reproduces fairly well the
data, both concerning absolute values and slopes of the correlations.
Again, it is the first time (to our knowledge) that a detailed comparison of this
type of data is made to fully self-consistent models of galactic chemical
and photometric evolution. We think that the results of this comparison 
are quite satisfactory.

\begin{figure*}
\psfig{file=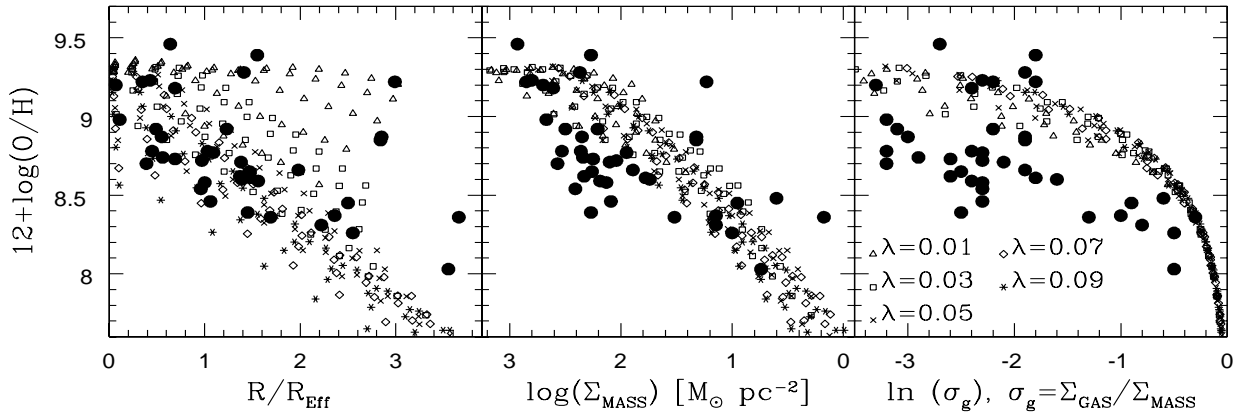,height=12.cm,width=\textwidth,angle=-90}
\vspace*{-5.5cm}
\caption{\label{SHIELDS}  
Oxygen abundances as a function of radius (in units 
of the effective radius R$_{Eff}$,
{\it left pannel}), local surface density ({\it middle pannel}), and
local gas fraction ({\it right pannel}).
Our model results ({\it open sumbols}) are given at an age of 13.5 Gyr
and correspond to different galactocentric distances; they are parametrised
by the corresponding $\lambda$ value, as indicated on the right pannel.
Data ({\it filled symbols}) are from Shields et al. (1991) 
for HII regions in field spirals.
}
\end{figure*}

Zaritsky et al. (1994) point out that the trend of $Z$ vs. M$_B$ observed
in their galactic disks is simply the continuation of the abundance-luminosity
relation noticed by Garnett and Shields (1987) over a wide range of 
magnitudes between spirals and irregulars. The relation spans over 10
magnitudes in M$_B$ and over a factor of 100 in metallicity. If galactic mass 
is always driving this relation (as in our models), then little 
room is left for galactic winds to play an important role; indeed, it 
is often claimed that low mass galaxies suffer loss of metals and/or gass
through galactic winds, but in that case a deviation from the correlation
observed in high mass galaxies (presumably unaffected by mass loss)
would be expected.

It should be noticed that Zaritsky et al. (1994) find no correlation
between metallicity and gas fraction $\sigma_g$ in a limited sub-sample 
(14 galaxies) of their data. As shown in Fig. 8, we do find such a correlation
in our results (higher $Z$ for lower $\sigma_g$). However, this correlation
is stronger for high $\sigma_g$, while at low $\sigma_g$ (below $\sim$0.2)
metallicity tends to ``saturate''. Zaritsky et al. (1994) admit that their
negative result may be due to the fact that their sub-sample is limited to
relatively low gas fractions.
It is interesting to notice that our results show also that metallicity tends to
saturate at a value of log(O/H)+12 $\sim$ 9.3 for the most massive disks, i.e.
those with the lowest gas fractions according to Fig. 8.

The results of Zaritsky et al. (1994) concern abundances as a function of
{\it global} galactic properties. Abundances in spiral galaxies as a function
of {\it local} properties were studied by Shields et al. (1991), who
searched for environmental effects between field and cluster galaxies (those
in Virgo). They compiled data on abundances in HII regions and atomic and 
molecular gas  surface densities as a function of galactocentric radius in 
five field galaxies. We concentrate      on those field galaxies only, that
presumably evolved ``passively'' (i.e. with no environmental effects) and
correspond better to our models.

The data of Shields et al. (1991) are plotted in Fig. 15 and compared to our
results. Oxygen abundances are plotted as a function of radius $R$ (normalised to
effective radius $R_{Eff}$), of total surface density $\Sigma$ and of gas
fraction $\sigma_g$. Our results are shown at an age of 13.5 Gyr, parametrised
by the values of spin parameter $\lambda$. 
It can be seen that: 

1) There is a good anti-correlation in general between metallicity and 
$R/R_{Eff}, \Sigma$ and $\sigma_g$ ({\it left, middle} and {\it right}
pannels, respectively). The scatter in the observed relation is
larger in the case of $Z$ vs. $R/R_{Eff}$ and smaller in the case of $Z$ vs.
$\Sigma$.

2) Our models reproduce well, in general, the observed trends. However, the 
scatter in the observed $Z$ vs. $R/R_{Eff}$ relation is only reproduced
when $\lambda$=0.01 models are considered: they are the only ones producing 
high metallicities at large  $R/R_{Eff}$. But such $\lambda$ values produce
 unrealistic disks, as stressed several times in this work ; and if we
neglect them, we are unable to reproduce the observed scatter. The solution 
to that difficulty may reside in the fact that the large abundances at large
$R/R_{Eff}$ in the Shields et al. (1991) data come from only one galaxy, N2903.
If this, perhaps ``peculiar'', galaxy is removed from the sample, the rest of the
data present a tighter  correlation that is readily explained by our models
with $\lambda >$0.02.

3) The metallicity vs. surface density relation presents the tightest 
correlation, both observationally and in our models. We find an excellent
agreement between theory and observations in that case.

\begin{figure*}
\psfig{file=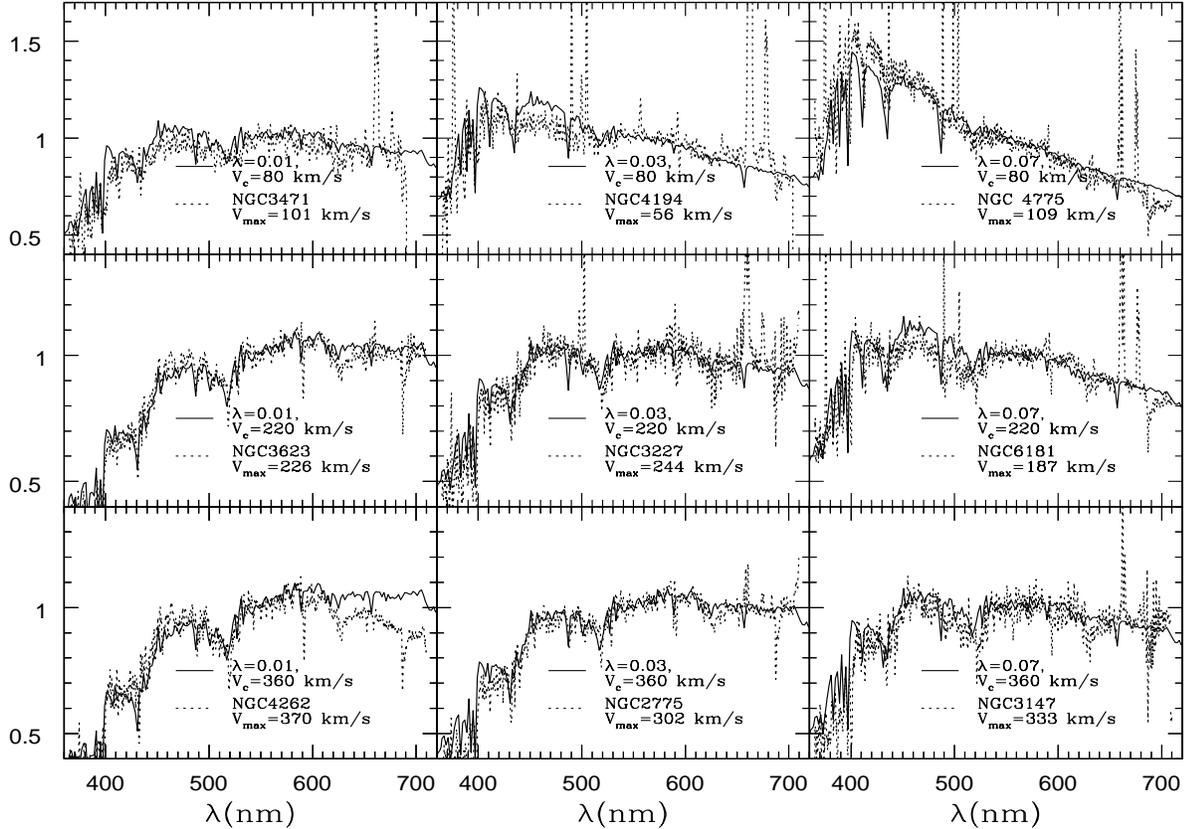,height=12.cm,width=\textwidth,angle=-90}
%\vspace*{-5.5cm}
\caption{\label{SPECTRES}  
Spectra of our model disks at an age of T=13.5 Gyr, compared to
observed spectra from the Catalogue of Kennicutt (1992).
We have selected a few representative values of $\lambda$ (0.01, 0.03, 0.07
from right to left) and of $V_C$ (80, 220, 370 km/s, from top to bottom)
and plotted the corresponding integrated galactic spectra at T=13.5 Gyr
({\it solid curves}).
For each of these cases, we selected in Kennicutt's catalogue
spectra of galaxies with measured  rotational velocities (V$_{max}$ in
each pannel, following the galaxy's name) as close as possible 
to the model values; they are plotted  in {\it dotted curves}.
All spectra are normalised to the flux at 550 nm.
Notice that our model spectra do not include emission lines.
}
\end{figure*}

4) The metallicity vs. gas fraction relation is the most intriguing. Our model
results are systematically $\sim$0.2-0.3 dex higher than the observations.
One might think that the adopted yields are too high by $\sim$50\%; such an
uncertainty in stellar yields is indeed conceivable (see Prantzos 1998).
However, in that case our results as a function of $R/R_{Eff}$ and $\Sigma$
should also be modified:
lowering the theoretically predicted O/H by 0.2-0.3 dex would  spoil the 
excellent agreement with the
data. Another possibility is that gas fractions have been systematically
underestimated in Shields et al. (1991): moving the data points to the right
(in the {\it right pannel} of Fig. 15) could alleviate the discrepancy. Also,
a combination of those two factors (i.e lowering the theoretical metallicity 
and increasing the observed gas fraction) would help. We wish, however, to 
point out an interesting parallel with the situation in our own Galaxy. It has
been noticed on  several occasions (e.g. Pagel 1997) that solar CNO abundances are
higher by 30-50\% than those in nearby young stars and the local interstellar
medium (ISM). Standard galactic chemical evolution models are usually 
required to reproduce solar abundances 4.5 Gyr ago and they always find 
present day CNO abundances larger by $\sim$30-50\% than those observed
in the local ISM (while reproducing correctly the present-day local gas
fraction, see e.g. Prantzos and Aubert 1995). Our models being calibrated on
the Milky Way, it is interesting to see that they overproduce
also present day oxygen abundances in field spirals, and by similar amounts!
This constitutes one more 
indication that solar abundances may not be representative of the local
ISM 4.5 Gyr ago, as pointed out in several works (Prantzos et al. 1996, 
Clayton 1997).
Confirmation of this would have major implications for galactic chemical
evolution studies.

\subsection{ Spectral evolution: integrated spectra}

We calculated the integrated spectra of our galactic disk models, according
to the procedure presented in BP99. Our aim is to see whether our models
can match reasonably well observed spectra of spiral galaxies.

In Fig. 16 we show our results for three values of $\lambda$ (0.01, 0.03 and
0.07, from {\it left to right}) and three values of $V_C$ (80, 220 and 360 km/s,
from {\it top to bottom}) at an age of 13.5 Gyr ({\it solid curves}).	
 As cen be seen, the form of  the spectrum depends on both
$\lambda$ and $V_C$. Low values of $\lambda$ and high values of $V_C$ lead to 
old stellar populations and to a severe depression  of the spectrum 
``bluewards'' of 400 nm, due to an efficient early star formation (see Fig. 4 and
8).

We searched in Kennicutt's (1992) catalogue of galactic spectra galaxies with measured
rotational velocities close to our $V_C$ values. For each one of them we were
able to find observed spectra that match reasonably well our theoretical  ones
({\it dotted curves} in Fig. 16). This is the case, for instance, of the
large spirals NGC2775 and NGC3147 ($V_{max}$=302 and 333 km/s, respectively,
in the middle and righ bottom pannel of Fig. 15).

We stress that the successful comparison of observed spectra to our
models proves nothing about the quality of the latter. Indeed, there are 
several ways to obtain spectra corresponding to young, intermediate and
old stellar populations, and use of composite spectra alone is a poor
diagnostic tool (especially if extinction has to be taken into account).

However, the comparison of models to observations made in Fig. 16 (the first
ever made that involves measured rotational velocities), when combined
to other probes of galactic evolution (all those discussed in the previous
sections) constitutes  a powerful tool of galaxy diagnostics. Indeed, it 
would be interesting to see, for instance, what are the values of gas 
fraction, chemical abundances, colours, magnitudes, disk scalelengths etc.
for the observed galaxies of Fig. 16, and check whether they fit to the general
scheme elaborated in this work. Also, it would be interesting to find out
what makes galaxies with  similar $V_{max}$ (like NGC3471 and NGC 4775) display
so different spectra; our models suggest that spin parameter may be the answer.
In any case, a detailed study of this kind requires as uniform a set of data
as possible. Work is curently in progress in that direction (Boissier and 
Prantzos, in preparation).

\section{ Summary}
 
In this work, we present 
a comprehensive study of the chemical and spectrophotometric evolution of 
spiral galaxies, using multi-zone models. 
With respect to previous studies of that kind, our work presents the
following important features:

- The fully self-consistent treatment of chemical {\it and} spectrophotometric
  evolution, with up-to-date, metallicity dependent, stellar input data.

- The use of a realistic stellar IMF, i.e. {\it measured}
  in the solar neighborhood (Kroupa et al. 1993), instead of the often used
  Salpeter IMF in the whole mass range;
  also, the radial dependence of the SFR (proportional to the rotation
  frequency $V(R)/R$, with realistic $V(R)$ profiles), based on theories of star
  formation induced by spiral waves in disks.
  
- The calibration of the model on the Milky Way disk (BP99), fixing the star
  formation efficiency which is {\it not a free parameter} for the other 
  disk models. 

- The use of simple ``scaling laws'' for the properties (sizes and surface 
  densities) of 
  other galactic disks, that are adopted as boundary conditions in our study;
  these scaling relations are based on currently popular models of galaxy
  formation, in the framework of Cold Dark Matter senarios and allow to describe
  disks in terms of two parameters: rotational velocity $V_C$ and spin parameter
  $\lambda$.

The main results of our models may be summarised as follows:

i) The final gas fraction $\sigma_g$, $M/L$ ratio, metallicity and colours
depend on both $\lambda$ and $V_C$. Disks of low $\lambda$ (compact) and/or high
 $V_C$ (massive) are formed earlier (in less than 6 Gyr) and are ``redder''
today than disks with other ($\lambda,V_C$) values.

ii) Some disk properties are found to be quasi-independent of $V_C$ (like
central surface brightness $\mu_{0}$, in agreement with observations), while
others are independent of $\lambda$ (like  e.g. current star formation rate $\Psi_0$,
absolute magnitude M$_B$  or colour-colour relation).

iii) High $\lambda$ values ($>$0.07) lead to Low Surface Brightness galaxies
(as already shown in previous works, e.g. Jimenez et al. 1998), while $\lambda$=0.01
leads to unrealistically small and bright disks.

iv) Due to the inside-out star formation scheme adopted here, profiles of
surface brightness and colours flatten in time.

Comparison of our results to a large body of observational data leads to the
following conclusions:

1) Disk size and central surface brightness are well reproduce by our models,
implying that the adopted scaling laws produce ``realistic'' boundary conditions;
only $\lambda$=0.01 values produce unrealistic results, ressembling more to
galactic bulges than to disks.

2) The Tully-Fischer relationship is ``built-in'' in this type of models
(i.e. through the scaling relations, which constitute the boundary  conditions
of the problem), but
realistic senarios for star formation lead to small but important variations:
in particular, our models predict a scatter increasing with decreacing circular
velocity (a trend suggested by the data of Giovanelli et al. 1997) and 
 a steeper slope in the I-band for low surface brightness galaxies. 
Also, we find that extinction by dust makes the TF relation flatter than in the
case of the stellar population alone. Most importantly, our models
naturally predict an increase in the slope of the TF relation with decreasing wavelength,
resulting from the fact that less massive galaxies are chemically and photometrically
``younger'' than more massive ones; this trend is indeed observed (Tully et al. 1998).

3) Colour-colour and colour-magnitude relations are well reproduced with the
adopted (Milky Way-type) star formation efficiency, provided
the observed disks are more than $\sim$10 Gyr old. More massive disks are ``redder''
on average than low mass ones. 

4) The relation of gas fraction to magnitude M$_B$, colour B-V and central surface
brightness $\mu_{B0}$ is well reproduced. The trends in the former two correlations
are driven by galaxy's mass ($V_C$), while in the latter by the  spin parameter
$\lambda$. Some of the observed low surface brightness galaxies require models
with higher $\lambda$ values than those studied here. Observed high surface
brightness galaxies with large gas fractions ($\sigma_g >$0.3) cannot be
explained by our models.

5) Metallicity is found to be well correlated to mass and luminosity, in fair
agreement to observations. Also, it is correlated to local environment (radius,
total surface density), again in agreement with observations.
Our models show higher metallicity than observed at a given mass fraction,
reminding of an analogous situation in the Milky Way disk. This ``problem'' may
represent another clue against the usual assumption that the solar abundances are
typical of the local ISM 4.5 Gyr ago.

6) Observed spectra of spirals with measured rotational velocities are well
reproduced by  our models, when comparable values of $V_C$ are considered.
The spin parameter $\lambda$ can help to explain  differences between galaxies
with the same rotational velocity.

The agreement of our models to such a large body of observational data is
impressive, taking into account the small number of free parameters.
Indeed, the only really ``free'' parameter is the dependence of the
infall timescale on galactic mass (Fig. 3), since all other ingredients
have been fixed already by the calibration of the model to the Milky Way disk. 
Although it does not constitute a proof of the validity of the model, this
success suggests nevertheless a coherent  overall picture for the evolution
of galactic disks, along the following lines:

-  Baryonic gaseous disks form in non-baryonic haloes, with properties
given by simple scaling relations of the Cold Dark Matter senario.

- Timescales for star formation in  the disks depend mainly on their mass:
massive disks form {\it earlier} the bulk of their stars and reach   
higher metallicities and B-V values than their low mass counterparts.
This picture seems to contradict the idea that low mass haloes (and baryonic
disks) form first. However, the timescale for assembling the baryonic gas
need not be the same as the star formation time scale: low mass galaxies may
assemble relatively rapidly, but require a very long time to turn their 
gas into stars. This point is crucial, because we found that it allows
to explain simultaneously  three important observed features:
the variation of the slope with wavelength band in the 
TF relation (Sec. 4.2, Fig. 11), the fact that more massive galaxies
are on average ``redder'' than low mass ones (Sec. 4.3) and 
the metallicity-luminosity relation of 
Zaritsky et al. (1994, Fig. 14). It is important to notice that, the
assumption of mass-dependent galactic winds could help explain the
third feature, but not the first one.

- Star formation in disks takes place inside-out, producing steep early gradients
in metallicity and colour profiles, that flatten at late times.

Several other predictions of this senario can be checked against observations.
In a forthcoming paper (Boissier and Prantzos, in preparation) we show that the
resulting abundance gradients are in agreement with observations. Moreover, the
senario predicts little evolution of massive disks at late times (since most
of the action takes place early on) and more important late evolution for 
low mass disks. The former point seems to be supported by observations of 
disk properties up to redshifts of $\sim$1 (Lilly et al. 1998).

\label{lastpage}

\end{document}